\documentclass[12pt]{article}
\usepackage{graphicx}
\usepackage{amsmath}
\usepackage{subfig}
\usepackage{slashed}
\usepackage{multirow}
\usepackage{booktabs}
\usepackage{color}

\def\ie{{\it i.e.}}
\def\cf{{\it c.f.}}
\def\eg{{\it e.g.}}

\def\tev{\,{\rm TeV}}
\def\gev{\,{\rm GeV}}
\def\mev{\,{\rm MeV}}
\def\kev{\,{\rm keV}}

\def\to{\rightarrow}
\def\ch{\ifmmode H^\pm \else $H^\pm$\fi}

\newskip\zatskip \zatskip=0pt plus0pt minus0pt
\def\matth{\mathsurround=0pt}
\def\lsim{\mathrel{\mathpalette\atversim<}}
\def\gsim{\mathrel{\mathpalette\atversim>}}
\def\atversim#1#2{\lower0.7ex\vbox{\baselineskip\zatskip\lineskip\zatskip
\lineskiplimit  0pt\ialign{$\matth#1\hfil##\hfil$\crcr#2\crcr\sim\crcr}}}

\newcommand{\ra}[1]{\renewcommand{\arraystretch}{#1}}

\begin{document} 

\begin{titlepage}
\rightline{\vbox{\halign{&#\hfil\cr
&SLAC-PUB-15197\cr
}}}
\vspace{1in}
\begin{center}

{{\Large\bf Bounds on Dark Matter Interactions with \\ Electroweak Gauge Bosons}
\footnote{Work supported by the Department of
Energy, Contract DE-AC02-76SF00515}\\}
\medskip
\medskip
\normalsize
{\large R.C. Cotta, J.L. Hewett, M.-P. Le and
  T.G. Rizzo{\footnote {e-mail: randoo, hewett, myphuong,
      rizzo@slac.stanford.edu\
}} \\
\vskip .6cm
SLAC National Accelerator Laboratory,  \\
2575 Sand Hill Rd, Menlo Park, CA 94025, USA\\}
\vskip .5cm

\end{center}
\vskip 0.8cm

\begin{abstract}

We investigate scenarios in which dark matter interacts with the Standard Model primarily through
 electroweak gauge bosons. We employ an effective field theory framework wherein the Standard Model
 and the dark matter particle are the only light states in order to derive model-independent bounds.
 Bounds on such interactions are derived from dark matter production by weak boson fusion at the LHC,
 indirect detection searches for the products of dark matter annihilation and from the measured
 invisible width of the $Z^0$. We find that limits on the UV scale, $\Lambda$, reach weak scale values
 for most operators and values of the dark matter mass, thus probing the most natural scenarios in
 the WIMP dark matter paradigm. Our bounds suggest that light dark matter ($m_{\chi}\lsim m_Z/2$
 or $m_{\chi}\lsim 100-200\gev$, depending on the operator) cannot interact \emph{only} with the
 electroweak gauge bosons of the Standard Model, but rather requires additional operator contributions
 or dark sector structure to avoid overclosing the universe.

\end{abstract}

\renewcommand{\thefootnote}{\arabic{footnote}}\end{titlepage}


\section{Introduction}
\label{sec:intro}
Evidence for a substantial particle dark matter component in our galaxy is by now quite convincing. The strength and
nature of non-gravitational DM interactions with the Standard Model (DM-SM interactions) are
unknown, but sensitivities of many currently operating experiments are at the level of predicted signals for many
well-studied DM models. Possibilities range in principle from just beyond current experimental bounds to levels that 
are impossible to probe directly, although the predominant weakly-interacting-massive-particle (WIMP) dark matter paradigm
naturally provides for the observed DM relic abundance, while predicting DM-SM interactions mediated via couplings of
approximately weak interaction strength to mediators with approximately weak scale masses. As is well-known, WIMP DM may
be observed in three possible ways: ($i$) directly, through it's interactions with nuclei (and/or electrons) in
underground detectors, ($ii$) indirectly, through it's self-annihilation into Standard Model (SM) particles in 
space and, lastly, ($iii$) through it's production at colliders such as the LHC, appearing as an excess of missing
transverse energy (MET).

The study of WIMP DM has been historically dominated by ``top-down''
studies based on new physics frameworks ($\eg$ supersymmetry, extra dimensions, etc.) that are primarily 
meant to solve the gauge hierarchy problem. Such theories naturally posit new particles with weak-interactions and
masses not far from the weak scale and more or less automatically provide WIMP DM candidates. In the current era much
 effort (rightly) continues to be devoted to such ``top-down'' studies ($\eg$, 
\cite{Cotta:2009zu}-\cite{CahillRowley:2012rv}), wherein
 the variety of phenomenology available in these frameworks can be studied in the context of fully-developed UV complete theories.

In quite an orthogonal direction a more ``bottom-up'' approach to studying DM-SM interactions has recently taken shape \cite{Beltran:2010ww}-\cite{Cheung:2012gi}. 
In the case where the DM particle is lighter than the degrees of freedom which mediate DM-SM interactions one can describe such 
interactions in an effective field theory (EFT) framework. In this picture the SM particles and the DM particle are the only 
light degrees of freedom in the theory and DM-SM interactions are described in terms of contact operators. 
At any fixed naive scaling dimension there are a limited number of such operators that respect Lorentz and gauge invariance so that, under
the assumption that one particular operator in such a set is the dominant interaction channel, one can systematically derive bounds in a
 fairly model-independent fashion. This approach lends itself particularly nicely to
studies of the complementarity of different classes of DM search experiments since, in the contact operator approximation, the very 
same coupling that dictates DM elastic scattering is, to some extent, also that which dictates DM pair production at colliders and
that which determines the energetic products of DM annihilation in astrophysical dark matter halos\footnote{In practice there are
  several caveats to this logic. For example, while the operator description may fairly generically
  apply to non-relativistic DM scattering and annihilation it may not be appropriate for typical LHC events with $\sqrt{\hat{s}}\gsim 1\tev$.
  It may also be the case that different operators are dominant in the different classes of experiments ($\eg$, a DM-quark
  interaction may dominate elastic scattering while a DM-lepton interaction may provide the dominant DM annihilation channel).}. This
approach has been used to great effect in the works \cite{Beltran:2010ww}\cite{Goodman:2010yf}\cite{Goodman:2010ku}\cite{Bai:2010hh}\cite{Rajaraman:2011wf}, in which it was
shown that if the EFT description holds, the irreducible collider signals provided by DM pair production with
associated initial state radiation (monojets) can place bounds on DM interactions with colored SM particles that are competitive
with dark matter direct detection experiments, especially for light ($m_{DM} \lsim 10 \gev$) dark matter. This approach has
since been developed in many directions, $\eg$, to include more general classes of operators \cite{Goodman:2010qn}\cite{Fox:2011fx}\cite{Rajaraman:2012db}
 and to include a wider variety of collider and DM searches \cite{Fox:2011pm}\cite{Fox:2012ee}\cite{Cheung:2012gi}\cite{Abazajian:2011tk}.

In this work we investigate the interactions between dark matter and the electroweak gauge bosons of the Standard Model. One reason 
to study this class of operators is simply that, ignorant as we are about DM-SM couplings, the most visible interactions (DM interactions
 with colored SM particles) may be suppressed relative to DM interactions with SM vector bosons. Besides this, such interactions
 may also indicate the extent to which DM is related to electroweak symmetry breaking 
and the extent to which the WIMP paradigm holds\footnote{Of course the non-observation of DM-gauge boson interactions cannot negate the WIMP
 picture by itself, as the ``WIMP miracle'' is actually in effect in far more diverse scenarios
 \cite{Feng:2008ya}\cite{Kumar:2009bw}.}. In ``top-down'' models the WIMP DM candidates are typically new partners of the electroweak gauge
bosons (such as the gauginos in SUSY and the KK photon of universal extra-dimensions models) or even of the higgs
 sector particles themselves (higgsinos in SUSY). As such this class of operators may provide information that
 is highly complementary to studies of WIMP dark matter in the context of UV complete theories of dark matter.

To begin we will describe our effective operator description of DM interactions with electroweak gauge bosons in detail (Section \ref{sec:operators}), focusing
on fermionic DM and considering operators of naive scaling dimension $d\leq 7$. To set bounds on DM interactions with electroweak gauge bosons we consider
weak boson fusion (WBF) searches for production of dark matter in $8\tev$ and $14\tev$ LHC searches (Section \ref{sec:wbf}). We then consider DM indirect-detection
bounds that can be obtained from current observations of dwarf spheroidal galaxies, monochromatic $\gamma$-ray searches and from measurements of the cosmic-ray
antiproton spectrum (Section \ref{sec:dm}). Finally we combine collider and indirect-detection limits and discuss our results. 
As will be seen below the combined reach of these searches is quite significant, $\sim 1$ TeV, for a wide range of $\chi$ masses. The discussion presented
here is related to the works \cite{Rajaraman:2012db}\cite{Frandsen:2012db}\cite{Bell:2012rg}\cite{Weiner:2012cb}\cite{Weiner:2012gm}.

\section{Dark Matter Effective Theory}
\label{sec:operators}

 In this section we first present a general discussion of the philosophy and structure of our EFT model of DM interactions with electroweak
gauge bosons, which leads us to a list of operators that will be the main focus of this paper (Table \ref{optable}) and their detailed description.

We assume that the relevant light degrees of freedom include only the usual matter content of the SM and a dark matter particle $\chi$. 
Here we consider only Dirac or Majorana fermionic\footnote{In this document we will use the symbol $\chi$ to refer to both the DM
 particle itself and to the multiplet containing the DM particle. When it is not clear from the context we further clarify using
 $\chi^0$ and $\chi^{\pm}$.} $\chi$, though generalization to scalar $\chi$'s would be a straightforward extension
of this analysis. We consider operators of naive scaling dimension $d\leq 7$ coupling $\chi$ to the SM bosons $\gamma$, $Z^0$ and $W^{\pm}$
 (or equivalently $B^0$ and $W^a$), which we may generically refer to as $V$. In general we have contact operators not only of the 4-point
 $\chi\chi VV$ topology, but also
operators which generate 3-point $\chi\chi V$ couplings to neutral gauge bosons, the latter of which (as we will discuss at length in
the following sections) lead to substantial alterations to our phenomenological expectations. We require all operators
to satisfy $U(1)_{EM}$ invariance, but not necessarily invariance under the unified electroweak $SU(2)_L\otimes U(1)_Y$.
This allows for the possibility that our effective theory is UV completed by a theory which has already undergone
electroweak symmetry breaking, so that operators such as the $d=5$ Higgs portal \cite{Patt:2006fw}-\cite{Kanemura:2010sh} operator $\bar{\chi}\chi V_{\mu}V^{\mu}$
 (where $VV$ can be $Z^0Z^0$ or $W^+W^-$) may be considered here. Throughout this work we present bounds assuming Dirac $\chi$, as
corresponding bounds for Majorana $\chi$ differ only by the appropriate symmetry factors or vanish identically, as in the case
 of operators containing $\bar{\chi}\gamma^{\mu}\chi$, or $\bar{\chi}\sigma^{\mu\nu}\chi$.

Since we are not model-building UV complete dark sectors, but rather working in an effective operator approximation our assumptions 
about the electroweak charges and representations of the $\chi$ multiplet have little qualitative impact on our end results. Nevertheless, 
it is conceptually helpful to discuss this here in some detail. DM must be the electrically neutral and stable component of some multiplet
 $\chi$ and may be either a SM singlet or non-singlet, except in the case of operators such as the ``magnetic moment,'' 
$\bar{\chi}\sigma_{\mu\nu}t^a\chi V^{a\mu\nu}$, which obviously require a $\chi$ transforming under $SU(2)_L\otimes U(1)_Y$.
For multiplets having both charged and neutral components the charged states are generically heavier than the neutral
component by $\mathcal{O}(100\mev-1\gev)$ due to loop corrections \cite{Cirelli:2005uq}\cite{Cirelli:2009uv}, although a full description of the UV theory
 would be necessary to actually calculate this splitting and to assure that $\chi^0$ is the lightest new particle. Additional discrete symmetries ($\eg$,
 R-parity, T-parity, KK-parity, etc.) are typically posited to prevent $\chi^0$ from decaying into lighter SM states.
 Non-singlet $\chi$'s may be in either chiral or non-chiral representations of $SU(2)_L\otimes U(1)_Y$. DM in chiral 
representations must get its mass, 
like the SM fermions, via the Higgs mechanism and thus has a renormalizable coupling, $\chi\chi h^0$, to the Higgs. Such a coupling is 
not considered in the current analysis and could significantly modify our results (although models with additional chiral matter 
are highly constrained by precision observables). Non-chiral $\chi$'s may be further classified as being in real 
or complex representations of $SU(2)_L\otimes U(1)_Y$ (see \cite{Essig:2007az}). For real representations $T^3=Y=0$, implying that the leading $\chi\chi VV$
interactions are t-channel $\chi^{\pm}$ exchanges built out of vertices $\chi^0\chi^{\pm}W^{\mp}$ (the canonical example being
the SUSY wino). For complex representations $\chi$ has $Y\neq 0$ and thus couples to $Z^0$ at tree-level, a coupling which badly violates
DM direct detection bounds unless it is highly suppressed. This suppression is usually accomplished with two or more such multiplets, adding operators 
which mix gauge eigenstates with differing hypercharge such that the lightest neutral component hardly couples to the $Z^0$ \cite{Essig:2007az} (the canonical
 example being the SUSY higgsino). In the current work we try to remain agnostic as to the particular details of the dark sector and seek 
to present our bounds so that they can be interpreted in the context of any particular UV theory.

While the EFT formalism offers simplicity and model-independence, its disadvantage, relative to the study of
fully-defined models, is that it is not generally applicable. The crux of this issue is that when experiments probe our contact
 operators with energies $\sqrt{s_{*}}$ such that the SM and DM particle
 are no longer the only light degrees of freedom ($\ie$, $\sqrt{s_{*}}\sim \Lambda$ for operators with dimensionful coefficient $\Lambda^{-n}$)
 we expect our EFT description to break down and the details of the UV theory to become important. For direct detection experiments
 looking for DM scattering, in which characteristic energies are $\sqrt{s_*}\sim\mathcal{O}(10\kev)$, the effective theory should essentially always 
hold (as evidenced by the typical $\chi^0-\chi^{\pm}$ mass splittings discussed above). For indirect detection experiments
 that look for the annihilation products of highly non-relativistic ($\upsilon\sim10^{-3}$) halo DM, in which characteristic energies
 are $\sqrt{s_{*}} \sim 2 m_{\chi}$, the effective theory should hold as long as the mass of particles that mediate these interactions 
is $M\gsim 2m_{\chi}$, as we have basically assumed in stating that the SM and $\chi$ are the only light degrees of freedom. One caveat to this 
is in cases where the masses of new particles which mediate the interaction are $M<2m_{\chi}$ (``light mediators''). Although the EFT formalism may seem  
inappropriate in such a scenario, some studies have investigated taking this approach \cite{Bai:2010hh}\cite{Fox:2011fx}\cite{Fox:2011pm}. For
 collider experiments the efficacy of the operator formalism is not as automatic. In probing these operators
 via weak boson fusion at the LHC we expect energies on the order of the partonic CM energy $\sqrt{s_*}\sim\sqrt{\hat{s}}\sim\mathcal{O}(\tev)$.
 Strictly speaking then, bounds that are set in WBF searches must translate to bounds on mediator masses in excess of the average partonic CM energy
 in order to be interesting in this context. 

It is interesting to try and determine which portions of the $\Lambda$ vs.\ $m_{\chi}$ plane cannot have a perturbative UV completion. This has been done in 
previous works, for example those studying $\bar{\chi}\chi\bar{q}q$ interactions, by identifying the dimensionful coefficient of these operators with the
 couplings and mass of an imagined heavy mediator exchange, $1/\Lambda^2\sim g_1g_2/M^2$, imposing $M\gsim 2m_{\chi}$ and
 $g_1^2g_2^2\lsim (4\pi)^2$, thus defining a perturbativity boundary in the $\Lambda$ vs.\ $m_{\chi}$ plane. The situation is somewhat more
 complicated in the case of $\bar{\chi}\chi VV$ interactions. Consider, for example, the Higgs portal operator $\bar{\chi}\chi V_{\mu}V^{\mu}$,
 which is a $d=5$ operator with coefficient $1/\Lambda$. If one imagines that the corresponding UV completion involves an s-channel exchange
 of some new heavy scalar particle $S$ then the new vertex $SVV$ in the UV theory is dimensionful, having a coefficient $g_S \upsilon_S$
that may be associated to the electroweak vev $\upsilon\sim246\gev$, but could just as well involve dimensionful numbers generated via the details
 of the dark sector physics. Similarly,
the coefficient of the $d=7$ operator $\bar{\chi}\chi V^{\mu\nu}V_{\mu\nu}$ is ambiguously connected to the parameters of the underlying UV physics, as
such an operator typically describes a process that happens at loop level, having cross-sections which are complicated functions of the couplings and 
masses of particles in the loop ($\eg$, $\chi^0_1\chi^0_1\to\gamma\gamma$ in the MSSM). Given such complications we omit further discussion of UV 
perturbativity in this work. In any given UV theory this boundary can be straightforwardly computed and compared to the exclusion limits that will
be presented below.

Unitarity offers a UV-\emph{insensitive} criterion for determining the efficacy of our EFT description of WBF events at the LHC \cite{Shoemaker:2011vi}. As is
 familiar from, $\eg$, 
pion scattering, amplitudes in a low-energy effective description may violate S-matrix unitarity as the energy of these interactions is increased.
Conversely, at any given energy of interest we may typically consider increasing $\Lambda$ so that, above some value, our effective description would not violate 
unitarity for interactions at this energy. For $\Lambda$ below this value the apparent violation of unitarity signals the failure of our EFT description and
 suggests that some modification would be necessary ($\ie$, a proliferation of operators or the detailed dynamics of a particular UV completion) to correctly
 describe the physics. For events at the LHC we have the further complication that the energy flowing through our operator is distributed according to pdfs
 and kinematics. The most stringent use of unitarity on our EFT description would be to consider the worst case scenario, imagining the full machine
 center-of-mass energy flowing through our operator in order to set our ``unitarity bound'' in\footnote{Keep in mind that this is not an experimental bound
 on the value of the UV scale in our low-energy effective description of DM, but rather, a bound on the values of $\Lambda$ for which we can consider using
 our EFT.} the $\Lambda$ vs.\ $m_{\chi}$ plane. Typical WBF events, however, would have much less energy flowing
through our operator because of pdf supression. We could then simulate events, calculating the energy flow through our operator in each, and derive contours in the 
$\Lambda$ vs.\ $m_{\chi}$ plane such that, say, 99\% of events do not violate unitarity. The interpretation of this is that most of the events that we include
in our calculation of the experimental bound are well-described by our EFT language, only a small fraction are not (in an actual experiment these events might result in
the production of particles that mediate this interaction and thus may not even look like WBF events) so the bound we derive should be fairly accurate. In what
 follows we will display contours for which 99\%, 90\% and 50\% of events\footnote{We do this calculation semi-analytically (details later) so our WBF analysis cuts
 have not been applied to these ``events.'' The numbers 99\%, 90\% and 50\% are only strictly correct then for operators with $100\%$ signal acceptance.} appear
 not to violate unitarity in $8\tev$ and $14\tev$ WBF events.

Finally, while we will focus entirely on DM interactions with gauge bosons in this work, it is sensible to ask how the inclusion of the
 Higgs boson would affect our results. With SM singlet or vector-like DM, as is our focus, there is no possibility for
 renormalizable interactions coupling the $\chi$ to the Higgs before electroweak symmetry breaking. However, all combinations $\bar{\chi}\chi$,
 $\bar{\chi}\gamma^{\mu}\chi$, etc., are SM singlets and can be joined to operators like $|H|^2$, $H^{*}t^{a}H$ and $H^{*}D_{\mu}H$ giving
 non-renormalizable interactions between the $\chi$ and the Higgs. After electroweak symmetry breaking these operators lead to, $\eg$, $\chi\chi h$,
 $\chi\chi Vh$, $\chi\chi V$ and $\chi\chi VV$ interactions, the former two of which we are not considering in focusing on only operators
 of the form $\chi\chi V$ or $\chi\chi VV$. There are two main consequences of omitting 3-point and 4-point operators with Higgses: in leaving out $\chi\chi h$
 interactions we will not have spin-independent scattering via the Higgs at tree level in our EFT, and in leaving out $\chi\chi Vh$ interactions we will 
 not be able to quote gamma-ray line bounds for the $\gamma h$ final state. Operators of the form $\chi\chi V$ or $\chi\chi VV$ arising from Higgses in 
 the Lagrangian do not change our analysis qualitatively, but result in re-interpretations of these results. Taking $\bar{\chi}\chi |H|^2$ for example, 
 gives a $\chi\chi h$ coupling that (after integrating out the Higgs) would simply return what we are calling the Higgs portal operator:
 $\bar{\chi}\chi V_{\mu}V^{\mu}$. Can this scenario be discerned from one in which some other (heavier) scalar mediates this interaction? At LHC
 energies, strictly speaking, the light Higgs cannot be integrated out, as it will actually be
 produced on-shell, but (as is usually done in light mediator analyses) one could use the narrow width approximation in order to derive collider
 bounds on $\bar{\chi}\chi V_{\mu}V^{\mu}$ in either case. In considering direct detection bounds, the coupling
 to the light higgs $\chi\chi h$ could give significantly larger spin-independent scattering rates at a given $\Lambda$ as compared to an interaction mediated by a heavy
 particle that does not couple to fermions directly at that same $\Lambda$ ($\ie$, $\bar{\chi}\chi V_{\mu}V^{\mu}$ is the \emph{only} interaction in the EFT).
 Despite these being conceptually distinct scenarios, they likely cannot be distinguished in direct detection experiments and bounds on each are simply
 different interpretations of the same data. Combinations like $H^{*}D_{\mu}H$ and $(H^{*}D_{\mu}H)(H^{*}D^{\mu}H)$ also return $\chi\chi V$ and
 $\chi\chi VV$ interactions that, in the end, yield operators that we are already considering, $\eg$,
 \begin{eqnarray*}
 \varphi^{\dagger}D^{\mu}\varphi&\stackrel{\langle\phi\rangle}{\longrightarrow}&\frac{1}{2}
\left( \begin{array}{cc} 0 & \upsilon \end{array} \right)^* \left[ \partial^{\mu}-igW^{a\mu}\frac{\sigma^a}{2}-ig'YB^{\mu}\right]  \left( \begin{array}{c} 0 \\ \upsilon \end{array} \right) \quad \sim Z^{0\mu}.\\
\label{dhiggs}
 \end{eqnarray*}
\subsection{List of Operators}
\begin{table}\centering
\ra{1.3}
\begin{tabular}{@{}rcccccc@{}} \toprule
\phantom{abc}Name\phantom{abc} & Expression & Norm. & Vertices & Sub-Procs. & Ann.\  \\ \midrule
$dim=5$:\phantom{abcd}\\
D5a\phantom{abc} & $\bar{\chi}\chi V^{a\mu}V^a_{\mu}$ & $\Lambda^{-1}$ & 4pt & $ZZ\mathrm{,}WW$ & $\upsilon^2$ \\
D5b\phantom{abc} & $\bar{\chi}i\gamma_5\chi V^{a\mu}V^a_{\mu}$ & $\Lambda^{-1}$ & 4pt & $ZZ\mathrm{,}WW$ & $1$ \\
D5c\phantom{abc} & $\bar{\chi}\sigma_{\mu\nu}t^a\chi V^{a\mu\nu}$ & $\Lambda^{-1}$ & 3/4pt & $A\mathrm{,}Z\mathrm{,}WW$ & $1$ \\
D5d\phantom{abc} & $\bar{\chi}\sigma_{\mu\nu}t^a\chi \widetilde{V}^{a\mu\nu}$ & $\Lambda^{-1}$ & 3/4pt & $A\mathrm{,}Z\mathrm{,}WW$ & $1$ ($VV$), $\upsilon^2$ ($f\bar{f}$) \\
$dim=6$:\phantom{abcd}\\
D6a\phantom{abc} & $\bar{\chi}\gamma_{\mu}t^a D_{\nu}\chi V^{a\mu\nu}$ & $\Lambda^{-2}$ & 3/4pt & $A\mathrm{,}Z\mathrm{,}WW$ & $1$ \\
D6b\phantom{abc} & $\bar{\chi}\gamma_{\mu}\gamma_5t^a D_{\nu}\chi V^{a\mu\nu}$ & $\Lambda^{-2}$ & 3/4pt & $A\mathrm{,}Z\mathrm{,}WW$ & $1$ ($VV$), $\upsilon^2$ ($f\bar{f}$) \\
$dim=7$:\phantom{abcd}\\
D7a\phantom{abc} & $\bar{\chi}\chi V^{\mu\nu}V_{\mu\nu}$ & $\Lambda^{-3}$ & 4pt & $AA\mathrm{,}AZ\mathrm{,}ZZ\mathrm{,}WW$ & $\upsilon^2$ \\
D7b\phantom{abc} & $\bar{\chi}i\gamma_5\chi V^{\mu\nu}V_{\mu\nu}$ & $\Lambda^{-3}$ & 4pt & $AA\mathrm{,}AZ\mathrm{,}ZZ\mathrm{,}WW$ & $1$ \\
D7c\phantom{abc} & $\bar{\chi}\chi V^{\mu\nu}\widetilde{V}_{\mu\nu}$ & $\Lambda^{-3}$ & 4pt & $AA\mathrm{,}AZ\mathrm{,}ZZ\mathrm{,}WW$ & $\upsilon^2$ \\
D7d\phantom{abc} & $\bar{\chi}i\gamma_5\chi V^{\mu\nu}\widetilde{V}_{\mu\nu}$ & $\Lambda^{-3}$ & 4pt & $AA\mathrm{,}AZ\mathrm{,}ZZ\mathrm{,}WW$ & $1$ \\\bottomrule
\end{tabular}
\caption{Table describing operators used in this work. A full description is given in the text.
}  
\label{optable}
\end{table}
 We now discuss the particular operators that we will be working with. Table \ref{optable} lists these operators, according
 to their naive scaling dimensions, along with some of their properties. Column-by-column in Table \ref{optable}, we have
 listed operator \emph{names}, Lagrangian \emph{expressions}, our choice of canonical
\emph{normalization}, available \emph{vertex} topologies in our EFT description of the operator, allowed 
\emph{sub-processes} and the scaling of the leading terms in the non-relativistic expansion of the analytic formulae describing DM \emph{annihilation}.

The $d=5$ operators D5a-b could result from exchanges mediated by new heavy scalar 
or pseudoscalar bosons. These operators necessarily require spontaneous breaking of $SU(2)_L\otimes U(1)_Y$ down to $U(1)_{EM}$ and consequently
 only the $Z^0Z^0$ and $W^+W^-$ subprocesses are allowed for these operators. The $d=5$ operators D5c-d are similar to dark magnetic and
 electric dipole moments\footnote{Note that, due to the identity $\gamma_5\sigma^{\mu\nu}=(i/2)\epsilon^{\mu\nu\alpha\beta}\sigma_{\alpha\beta}$,
  the operator D5d is equivalent to an operator of the form $\bar{\chi}\gamma_5\sigma_{\mu\nu}\chi W^{\mu\nu}$.} (respectively), which have been the subject
of much recent study \cite{Bagnasco:1993st}-\cite{Barger:2012pf}. As these operators have one field strength $V^{\mu\nu}$, they give rise to both 4-point
 $\chi\chi VV$ and 3-point $\chi\chi V$ contact interactions. Since we are talking about interactions involving two \emph{neutral} DM particles, the
 $V^{\mu\nu}$ in the D5c-d operators can be either $W^{3\mu\nu}$ or $B^{\mu\nu}$, giving specifically the subprocesses: $\chi\chi W^+W^-$, 
$\chi\chi Z^0$ and $\chi\chi A^0$. The $d=6$ operators D6a-b \cite{Buchmuller:1985jz} could arise
 via exchange of new neutral vector bosons ($\eg$, a $Z'$) and, since $V^{\mu\nu}=W^{3\mu\nu}$ or $B^{\mu\nu}$, give rise to the same 4-point and 3-point
 interactions as in the D5c-d case. Finally, The $d=7$ operators D7a-d typically arise from 1-loop diagrams ($\eg$, the
 $\tilde{\chi}_1^0\tilde{\chi}_1^0\to\gamma\gamma$ process in SUSY), and may occur through any of the $W^+W^-$, $Z^0Z^0$,
 $Z^0\gamma$ and $\gamma\gamma$ subprocesses.

 The operator expressions listed in Table \ref{optable} should all be understood as the sum of the expression listed and its complex
 conjugate expression. We employ canonical normalizations listed in the ``Norm.'' column of the table.

 In the ``Vertices'' column we distinguish operators that have \emph{only} 4-point interactions, from operators
 that have \emph{both} 3-point and 4-point interactions. This is distinction is very important, as we will describe in the following
 sections, for both the LHC WBF bounds and for direct detection bounds. For WBF, operators with both 3-
 and 4-point interactions tend to generate events which look more like the background $W/Z+jj$ processes than otherwise, and thus are
 somewhat harder to constrain at the LHC. As concerns direct detection, operators with only 4-point interactions must scatter via higher-order
 processes and current bounds have been estimated to be somewhat far from current experimental sensitivities 
 \cite{Essig:2007az}\cite{Freytsis:2010ne}\cite{Frandsen:2012db}, while operators with both 3- and 4-point interactions may scatter off of nuclei via tree-level
 exchanges of the $Z^0$ or the $A^0$. Since we know that dark matter is \emph{dark} and approximately collisionless, the 3-point
 coupling $\chi\chi A^0$, and resulting long-range interaction, is highly constrained by an array of measurements 
 \cite{McDermott:2010pa} that are sensitive to both DM-SM interactions and to DM-DM self-interactions.
 The direct detection phenomenology of the $\chi\chi A^0$ scenario diverges from that which
 is usually studied, as the recoil spectrum derived for such a long-range interaction is distinct from the recoil spectra
 derived for the typically assumed $\bar{\chi}\chi\bar{N}N$ (SI) and $\bar{\chi}\gamma^{\mu}\gamma_5\chi\bar{N}\gamma_{\mu}\gamma_5 N$ (SD) interactions, and 
 has been the subject of recent developments \cite{Feldstein:2009tr}\cite{Fan:2010gt}\cite{Fitzpatrick:2012ix}. Given that such a scenario is so tightly
 constrained we will suppose that the $\chi\chi A^0$ interaction is negligible in setting our bounds, $\eg$, by tuning Lagrangian terms involving the
 $W^a_{\mu\nu}$ and $B_{\mu\nu}$ fields to cancel this vertex. The only opportunity for appreciable scattering rates is thus the $\chi\chi Z^0$ vertex.
 For our operators, however, these vertices always involve derivatives and are thus momentum suppressed and always result in negligible scattering.
 For $m_{\chi}\lsim m_Z/2$ the $\chi\chi Z^0$ vertex will allow the decay $Z^0\to\chi\chi$, contributing to the invisible width of the $Z^0$, which
 is measured to be $\Gamma_{Z\mathrm{,inv.}}\lsim 2\mev$ \cite{Nakamura:2010zzi}. The constraint from this bound will be shown in the figures that follow.

 The entries in the ``Sub-Procs.'' column refer to the allowed combinations of gauge bosons ($W^+W^-$, $Z^0Z^0$, $\gamma Z^0$ and $\gamma\gamma$ - 
 ``\emph{subprocesses}'') that can arise from the generic $V^{\mu}$ and $V^{\mu\nu}$ in each operator. Available subprocesses are determined
 simply by $U(1)_{EM}$ invariance, as vertices need to conserve electric charge and photons are only allowed to arise from the field strength $F^{\mu\nu}$.
 Any UV complete theory will specify the exact weighted combination of subprocesses present in the EFT, but here we will need to make some assumption about
 these combinations in order to proceed. As is discussed in the following sections this assumption is more subtle in the context of
 WBF, where the gauge bosons are intermediate state particles and different subprocess combinations correspond to different 
 \emph{coherent} sums. One may also expect that custodial symmetry \cite{Sikivie:1980hm}\cite{Nakamura:2010zzi} is an important constraint on the allowed
 subprocess combinations. We will
 not consider this in further detail, except to say that corrections ($\eg$, to gauge boson masses) are suppressed both by the appropriate power of 
 the high scale $\Lambda$ and also by an additional loop factor, so that combinations which deviate from the custodial limit are not obviously ruled out.
 In the end it is likely necessary that corrections to electroweak precision observables need to be calculated in the complete UV theory.

 The column labeled ``Ann.'' lists the leading order terms in the non-relativistic expansion of the analytic calculation of DM annihilation.
 For operators D5d and D6b this is quoted in terms of annihilations to either the $VV$ or $f\bar{f}$ final states, as they have different 
 leading order terms in their expansions. We will obviously expect that operators with leading order terms $\sim\upsilon^2$ will be significantly more
 difficult to constrain via indirect detection searches than those with leading order terms $\sim\mathcal{O}(1)$. 

 Finally, while not shown on the table, we note that the various operators have various P and CP properties. While parity violation in the dark sector is
 unconstrained (and possibly motivated by the observed parity violation in the SM), CP violation in the dark sector is possibly constrained
 via the CP violation that could be induced in the SM via higher-order interactions with DM in loops. Such constraints are
 expected to be highly model-dependent and are thus beyond the scope of this work. 

\section{Bounds from Weak Boson Fusion at the LHC}
\label{sec:wbf}
 Weak boson fusion (WBF) processes have been widely studied
 as a means of enhancing LHC searches for the Higgs boson and such 
searches have been shown to be an effective strategy for discovering
invisibly decaying Higgs bosons\footnote{with cross-sections, after cuts, of
order of $\sim100\;fb$ at the $14\tev$ LHC}. Here we will use this 
same type of analysis, not for studying the Higgs, but for setting bounds
on our contact operators. One could also imagine using searches for a single
electroweak vector boson recoiling off of missing transverse momentum, as has
been discussed in the context of searches for an invisibly decaying Higgs in, 
$\eg$, \cite{Davoudiasl:2004aj}. Here we focus only on WBF, as the enhanced production
through longitudinally polarized gauge bosons has the potential for further reach than
otherwise. For operators that favor transversely polarized gauge bosons, however, we expect 
that such ``mono-V'' searches would complement the results derived here. WBF
signal events are characterized by the presence of two very energetic and
well-separated forward/backward jets, as well as large missing
transverse momentum. Here we derive bounds on our operators from WBF searches
analyses of (current) $8\tev$ $25\;\rm{fb}^{-1}$ and (future) $14\tev$ $100\;\rm{fb}^{-1}$
LHC data sets.

This section is divided up into three parts: first we discussed the application of 
the contact operator approach in the context of WBF searches, next we discuss the details
of our WBF analysis in detail, and finally we present the resulting bounds on our contact operators.

\subsection{Contact Operators and Weak Boson Fusion}
\label{sec:method}
As mentioned in the introduction, employing our EFT description in the 
collider environment is much subtler than an EFT description of direct and indirect
dark matter searches. The first issue that we need to address is the treatment of 
\emph{subprocesses} in our WBF analysis. Generic UV completions of our EFT will result in non-trivial
\emph{combinations} of contact interactions connecting the DM particle $\chi$ to the gauge boson pairs
$W^+W^-$, $Z^0Z^0$, $\gamma Z^0$ and $\gamma\gamma$ (\emph{subprocesses}). For indirect
detection searches different combinations of \emph{final} state gauge bosons will simply result in a different combinations of exclusion limits 
from the various DM searches. In a WBF search however, the gauge bosons are \emph{intermediate state} particles, and in principle the different subprocesses
add coherently. Here we have dealt with this difficulty as follows.

In a particular UV theory one can explicitly calculate the coefficients
that relate these different subprocesses, which we denote schematically
as:
\begin{eqnarray}
\alpha_{W^+W^-},~\alpha_{Z^0Z^0},~\alpha_{\gamma Z^0},~\mathrm{and}~\alpha_{\gamma\gamma}.
\end{eqnarray}
If only one of these $\alpha$'s is non-zero then we can obtain the exact bound by 
doing a collider simulation and calculating a cross-section for that particular subprocess operator in 
isolation\footnote{Here the $(s)$ is a reminder that these $\sigma$'s are dependent on the machine center of mass 
energy and also on the cuts employed in the analysis.},
\begin{eqnarray}
\sigma_{W^+W^-}(s),~\sigma_{Z^0Z^0}(s),~\sigma_{\gamma Z^0}(s),~\mathrm{or}~\sigma_{\gamma\gamma}(s),
\end{eqnarray}
 What we will do for this analysis is to simply 
present limits based on these individual subprocesses, and to estimate the limit on an operator involving
a sum of sub-processes by the weighted \emph{incoherent} sum of these limits:
\begin{eqnarray}
\sigma_{tot}\equiv\sum_i
\alpha_i\sigma_i(s)~~~~~~~\mathrm{where,}~~i \; \epsilon \;\{ W^+W^- \mathrm{,}~Z^0Z^0\mathrm{,}~\gamma
Z^0\mathrm{,}~\gamma\gamma \}.
\label{methodsum}
\end{eqnarray}
Of course, there should in principle be constructive or destructive \emph{interference}
amongst the amplitudes for different subprocesses, resulting in larger or smaller total cross-sections.
It is clear that the procedure described above cannot account for this, however we find
 that the error incurred in employing this approximation is typically small compared to the
 other sources of error ($\eg$, systematic uncertainties) present in the estimation of the bound.

To give a concrete example, suppose that the
operator D7a is the relevant WBF operator resulting from integrating
out the heavy fields in some particular UV theory. We then assume that the
$SU(2)_L$ and $U(1)_Y$ fields $W^{a\mu\nu}$ and $B^{\mu\nu}$ appear
with relative weights $a$ and $b$. In this
 case the momentum space Feynman rules (``F.R.'' below) derived in the mass-eigenstate basis are:
\begin{align*}
\frac{1}{\Lambda^3} \bar{\chi}\chi ( a W^{a\mu\nu}W^a_{\mu\nu} +
b B^{\mu\nu}B_{\mu\nu} )~~~~\stackrel{~~~F.R.~~~}{\longrightarrow}~~~~
&\frac{4a}{\Lambda^3} X^{\mu_1\mu_2},\tag{$W^+W^-$}\label{WW} \\
&\frac{4(a c^2_w + b s^2_w)}{\Lambda^3}
X^{\mu_1\mu_2},\tag{$Z^0Z^0$}\label{ZZ} \\
&\frac{4 (a s^2_w + b c^2_w)}{\Lambda^3}
X^{\mu_1\mu_2},\tag{$\gamma\gamma$}\label{GG} \\
&\frac{4c_ws_w(a - b)}{\Lambda^3}
X^{\mu_1\mu_2},\tag{$\gamma Z^0$}
\label{exampleexplicit}
\end{align*}
where $p_1$, $p_2$, $\mu_1$ and $\mu_2$ are the momenta and Lorentz indices of
gauge bosons, $s_w$ and $c_w$ are the sine and cosine of the Weinberg
angle, and $X^{\mu_1\mu_2} = p^{\mu_2}_1 p^{\mu_1}_2 - p_1 \cdot
p_2 \eta^{\mu_1\mu_2}$. Here we define the $\alpha_i$'s by taking the squared
prefactors from the above equation (we actually generate the cross-sections $\sigma_i$ using
$\Lambda=1\tev$, as is explained in Section \ref{SignalWBF}):
\begin{eqnarray}
\alpha_{W^+W^-}=16a^2\left(\frac{1\tev}{\Lambda}\right)^6 &\mathrm{,}&
\alpha_{Z^0Z^0}=16(ac_w^2+bs_w^2)^2\left(\frac{1\tev}{\Lambda}\right)^6, \\
\alpha_{\gamma \gamma}=16(as_w^2+bc_w^2)^2\left(\frac{1\tev}{\Lambda}\right)^6 &\mathrm{,}&
\alpha_{\gamma Z^0}=16(c_ws_w(a-b))^2\left(\frac{1\tev}{\Lambda}\right)^6. \nonumber
\end{eqnarray}
Thus we compute $\sigma_{tot}$ for this operator as:
\begin{eqnarray}
\label{D7sigmatot}
\sigma_{tot}&\equiv&\sum_i\alpha_i\sigma_i(s)\\\nonumber
&=& 16\left(\frac{1\tev}{\Lambda}\right)^6
\{(a)^2\sigma_{W^+W^-}(s)+(a c^2_w + b
s^2_w)^2\sigma_{Z^0Z^0}(s)\nonumber\\
&+& (a s^2_w + b c^2_w)^2\sigma_{\gamma\gamma}(s)+ (c_w
s_w(a - b))^2 \sigma_{\gamma Z^0}(s)\}.\nonumber
\end{eqnarray}
We will explore the bounds obtained for the example given above in
further detail in Section \ref{SignalWBF} (after we describe our
numerical analysis in detail), where we will pay 
particular attention to the influence of interference (or lack thereof)
and systematic uncertainties on these results.

\subsection{Selection for WBF at LHC: SM Background generation}
\label{BGWBF}
\begin{figure}[hbtp]
    \centering
    \includegraphics[width=0.80\textwidth]{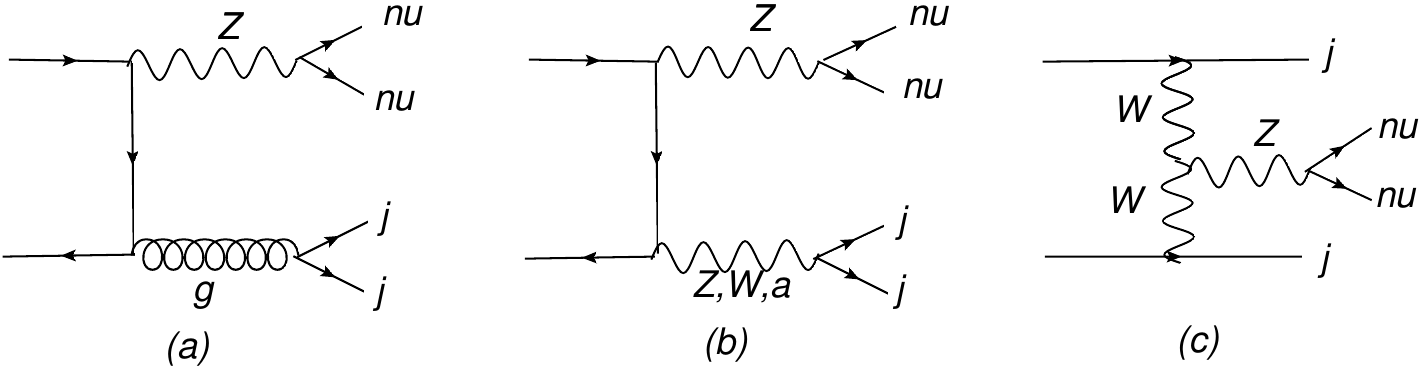}
    \caption{Example classes of feynman diagrams for $Zjj$ background, similar diagrams are found for $Wjj$ background as well.}
    \label{figs:BGfeyn}
  \end{figure}

 There are several classes of important backgrounds
to our signal, which are illustrated in Figure \ref{figs:BGfeyn}. We have: \emph{(i)} Drell-Yan processes with two
colored particles radiated off of the initial state serving as the two leading jets,
$\ie$, $Zjj$ and $Wjj$ where $Z^0\rightarrow\nu\bar{\nu}$ and $W\rightarrow l\nu$ (where in the latter case the lepton is not
identified), these are referred to as QCD Zjj or QCD Wjj backgrounds, \emph{(ii)} processes involving the t-channel exchange of a weak
gauge boson and further radiation of another EW gauge boson, these will be referred to as EW Zjj or EW Wjj and \emph{(iii)}
pure QCD backgrounds, where the mis-measurement of jets leads to
significant missing transverse momentum. Of these, the pure QCD
component can be substantially suppressed by cutting on $p^{miss}_T$, so
 we simply omit these events from our background
simulation (this was also found for the 3j simulations performed in the
work \cite{Eboli:2000ze}).

This analysis is carried out at the parton level, following the analysis
presented in \cite{Eboli:2000ze} as closely as possible. An important difference
between the two analyses is that, in the work \cite{Eboli:2000ze}, the authors
used various calculational tools to simulate the
various classes of events, while our analysis makes exclusive use of
the Madgraph v.5 \cite{Alwall:2011uj} package for all SM backgrounds,
as well as for the signal events arising from the operators described above.
While these differences are reflected in the cross-section limits derived in the two analyses we
observe that these discrepancies do not have a qualitative impact on setting limits
on the high mass scale, $\Lambda$.

\begin{figure}[hbtp]
    \centering
    \includegraphics[width=1.0\textwidth]{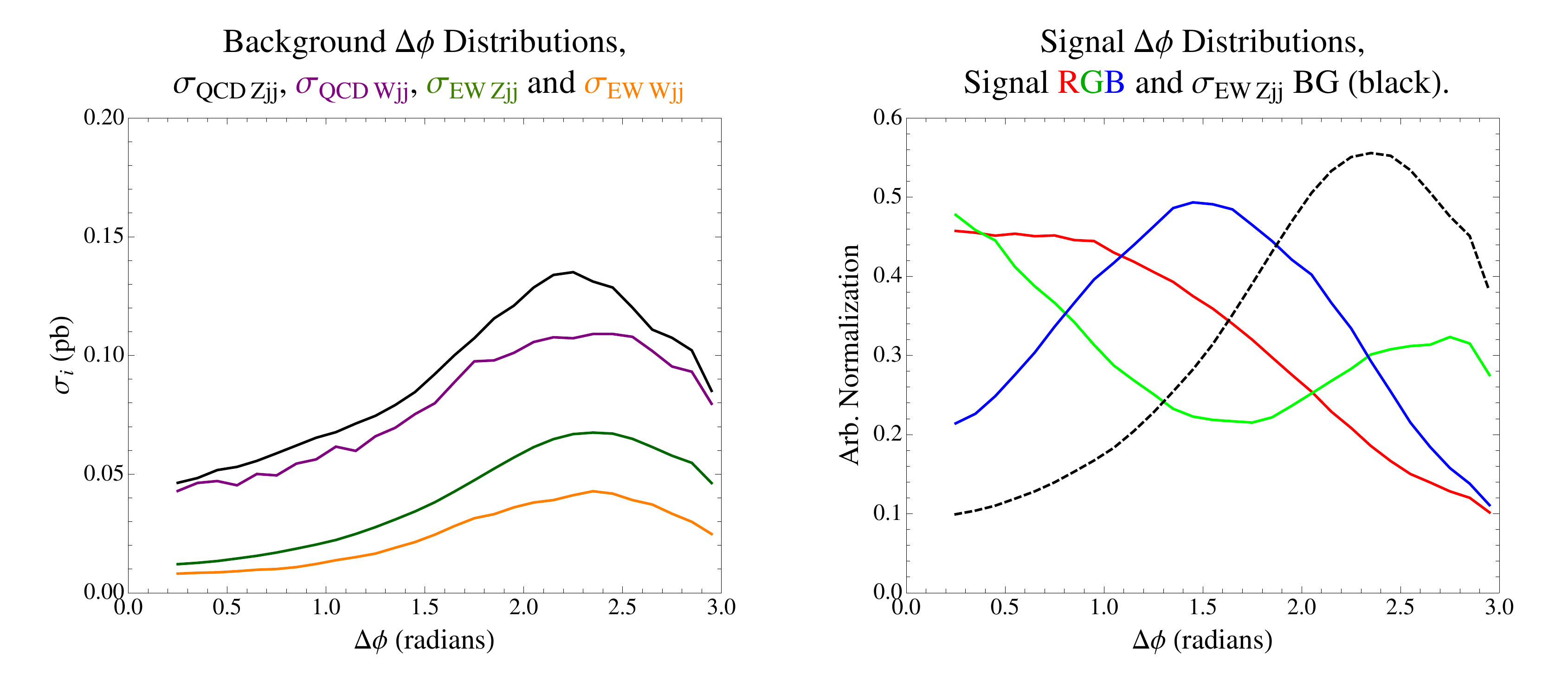}
    \caption{We display azimuthal interval ($\Delta\phi$)
    distributions for the background (left panel) and signals (right
    panel) used in this study. The background is split into
    components: QCD Zjj (black), QCD Wjj (purple), EW Zjj (green) and
    EW Wjj (orange). In the right panel curves are unit-normalized,
    the signal curves being described by red, green and blue curves,
    along with the background EW Zjj shape (black-dashed) for
    comparison. The operators D5a-D5d and D6a-D6b all correspond
    approximately to the red curve, while the operators D7a-D7b 
    correspond to the green curve and the operators D7c-D7d
    correspond to the blue curve.}
    \label{figs:azimuth}
  \end{figure}

Here we describe the most relevant cuts that have been applied in our
analysis and discuss their effects on the various background and signal components.
First, we impose cuts to select events with
hard forward and backward jets that are widely separated in pseudorapidity:
\begin{eqnarray}
p^j_{T} > 40GeV,~~|\eta_j| < 5.0,\nonumber \\ 
|\eta_{j1}-\eta_{j2}|>4.4,~~ \eta_{j1}\cdot\eta_{j2}<0.
\label{Jetcuts}
\end{eqnarray}
The origin of these cuts in the Higgs WBF analyses traces back to the fact that Higgses 
produced in weak boson fusion are produced predominantly by longitudinally polarized $W$ bosons
\footnote{This is more easily seen in the effective-W-approximation \cite{Cahn:1983ip}, where WBF is treated as a 2-to-2 
process initialized by two $W$'s, which are treated as approximately on-shell partons inside of the proton.}, 
which are radiated preferentially with $p_T\sim m_W/2$, giving high rapidity leading jets. In contrast, the 
jets produced in the background events are much more centrally distributed. Importantly, this latter fact is seen
to arise because of \emph{interference} between the two diagrams Figs.\ \ref{figs:BGfeyn}b and \ref{figs:BGfeyn}c.
As we will discuss in more detail later on, this has important implications for the \emph{signal} rates from
some of our operators as well.
  \begin{figure}[hbtp]
    \centering
    \includegraphics[width=0.80\textwidth]{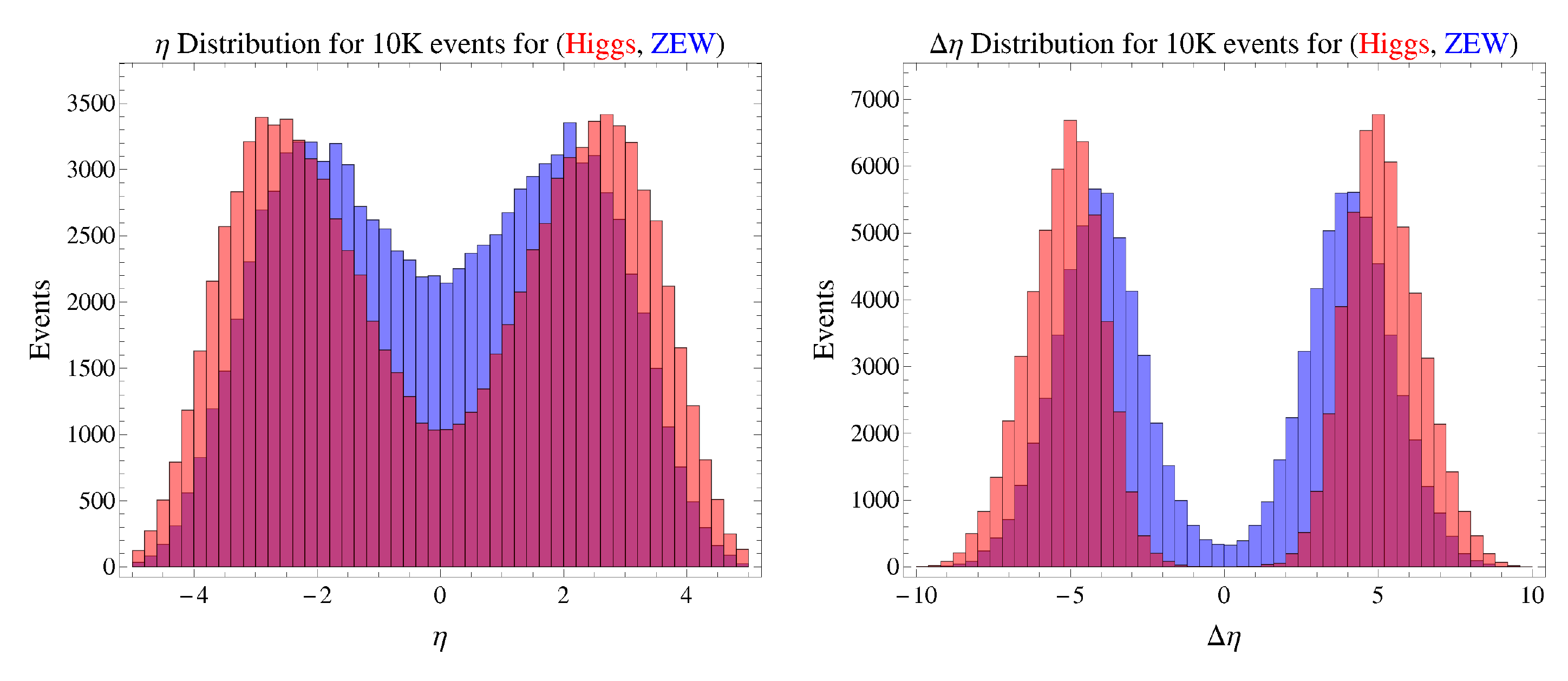}
    \caption{We display the jet rapidity distributions of a higgs-like signal (red) and electroweak Zjj background (blue)
      for samples of 10,000 events each (left panel.) Right panel shows the $\Delta\eta\equiv\eta_{j1}-\eta_{j2}$ distribution,
      we see that a cut of $\Delta\eta>4.4$ roughly contains only the tails of the background distribution while including
      the full peak of the higgs-like signal distribution. Other background components behave similarly.}
    \label{figs:etaHBG}
  \end{figure}
Next, we impose a cut on the missing transverse momentum of events,
$p^{miss}_T$, in order to remove the contributions from QCD 3j and the
soft single $\nu$'s resulting from W decays, as these contributions
fall off very quickly with energy above $\sim 100\gev$ ($\cf$, Fig.\ 1 of \cite{Eboli:2000ze}):
\begin{eqnarray}
p^{miss}_{T} > 100\gev.
\label{pTcuts}
\end{eqnarray}
We further impose a cut on the invariant mass of the two tagging jets
to suppress the contributions from QCD Zjj/Wjj background events,
whose radiated gluon jets are typically softer than those of the
corresponding quark jets in EW Zjj/Wjj (and signal) events ($\cf$, the
steeply falling $d\sigma/dM_{jj}$ of these events in Fig.\ 2 of \cite{Eboli:2000ze}):
\begin{eqnarray}
M_{j1,j2} > 1200\gev.
\label{Mcuts}
\end{eqnarray}
We further note that, since there is color exchange in the t-channel
of the QCD Zjj/Wjj processes, these events tend to result in higher
jet activity in the central part of the detector as compared to the
EW Zjj/Wjj and signal events. To account for this we follow
\cite{Eboli:2000ze}\cite{Chehime:1992ub} in simulating the effect of
requiring a $p_T > 20\gev$ veto on jets in the central region by
simply applying survival probabilities of 0.28 and 0.82 to QCD Zjj/Wjj
 processes and EW Zjj/Wjj processes, respectively, rather than
 actually applying the cut on an event-by-event basis. To remove a majority of Wjj background events, we veto events in which
we can confidently identify the lepton according to the criteria:
$|\eta_l| <$ 2.5 and $p_{Tl} > 5\rm{,}~10~\rm{and}~20\gev$ for e,
 $\mu$ and $\tau$ leptons, respectively. Finally, we apply a cut on the azimuthal interval, $\Delta\phi$,
 between the two tagging jets, which is especially helpful in
 discriminating events according to their Lorentz tensor structure. We
 apply the cut
\begin{eqnarray}
\Delta\phi=|\phi_{j1}-\phi_{j2}| < 1.
\label{phicuts}
\end{eqnarray}
It has been observed \cite{Eboli:2000ze}\cite{Hagiwara:2009wt}\cite{Buckley:2010jv} that such a cut favors Higgs-type
contact interactions, whose jets tend to be relatively close in azimuthal angle,
relative to the QCD Zjj/Wjj and EW Zjj/Wjj backgrounds whose jets tend
to be more back-to-back in the azimuthal plane. In Figure
 \ref{figs:azimuth} we give $\Delta\phi$ distributions for the various
 background components in this analysis, as well as for the various
 contact interactions used in this analysis. Since we are
 interested in signals coming from contact
 interactions with a variety of different Lorentz structures, ranging
 from the Higgs-like operators of the form D5a-b to
 $Z^{\prime}$-like operators of the form D6a-b, we don't always
 benefit from the discriminating power of this cut. Nevertheless, we choose to
 apply this cut in all cases in order to facilitate
 comparison with the existing literature \cite{Eboli:2000ze}. In any case, the effect of this cut on the
 resulting bounds is found to be relatively small.

\begin{figure}[hbtp]
  \centering
  \includegraphics[width=1.0\textwidth]{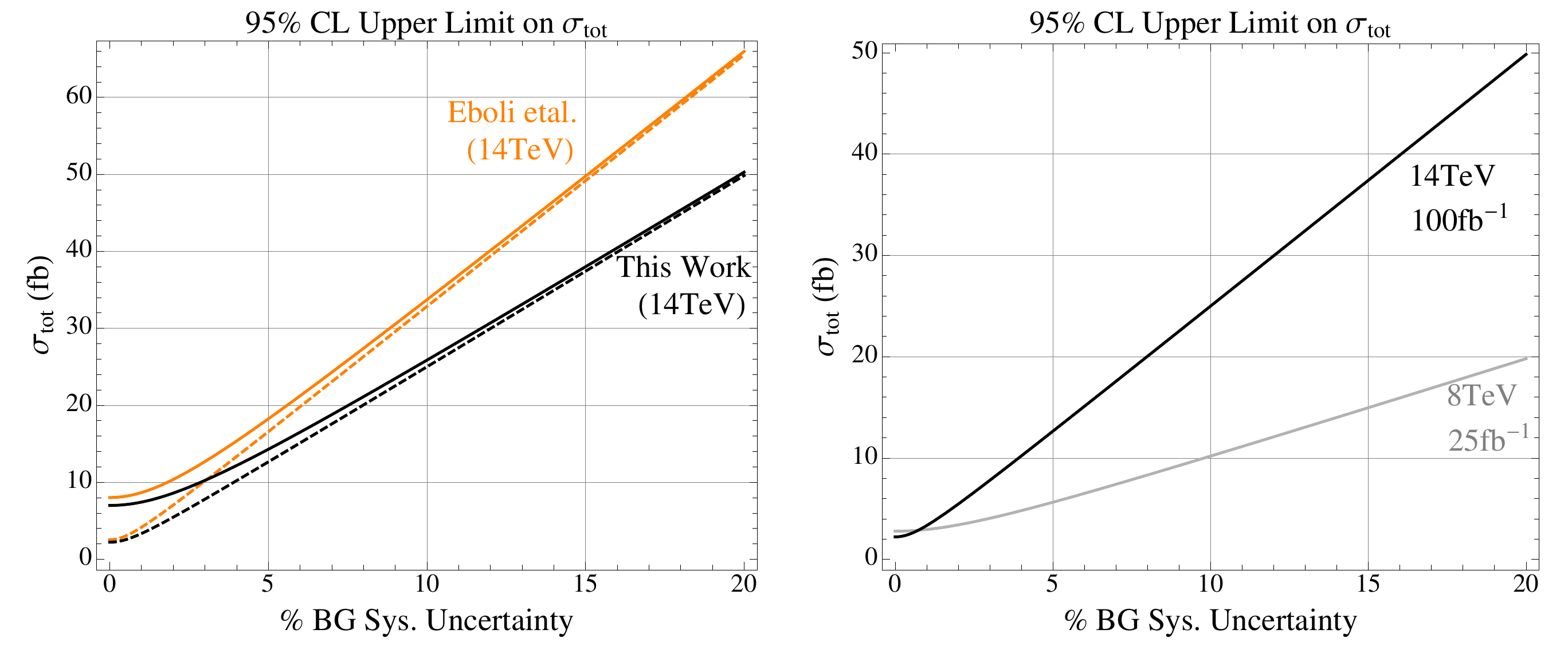}
  \caption{Here we display the 95$\%$ CL upper limit on the total signal
    cross-section, $\sigma_{tot}$ (fb) as a
    function of an assumed level of systematic uncertainty in the
    background. In the left panel we compare the results found in this work
    and those that would be found using the backgrounds derived in the work
    \cite{Eboli:2000ze} for $14\tev$ WBF analyses with $100\rm{fb}^{-1}$ (solid)
    or $10\rm{fb}^{-1}$ (dashed) data sets. In the right panel we compare our limits for a
    $14\tev$ $100\rm{fb}^{-1}$ analysis with those for a $8\tev$ $25\rm{fb}^{-1}$ analysis.}
    \label{limitsigma}
\end{figure}

 For comparison purposes, Table \ref{tableBG} shows the $14 \tev$ cross-sections
for the various background components generated in our study, along with those
found in \cite{Eboli:2000ze}.  We observe that our simulated QCD
Zjj/Wjj rates are about 70-80$\%$ of the size of those found in
\cite{Eboli:2000ze}, while the EW Zjj/Wjj rates from the two studies
match well. The authors of \cite{Eboli:2000ze} argue, using partially
data-driven background estimates, that one can achieve a combined
systematic uncertainty on the background in this analysis of
3.0(1.2)\;$\%$ for 10(100)\;$fb^{-1}$ at a $14\tev$ LHC.  These estimates
seem to be optimistic however, and we opt to instead quote 95\%
C.L. upper limits on total signal cross section, $\sigma_{tot}$, as a
function of the systematic
uncertainty on the backgrounds in Figure \ref{limitsigma}. In this way the reader can
estimate the power of such searches under any plausible assumption of
what the systematic uncertainty should be.

Results for $8\tev$ 25\;$fb^{-1}$ WBF analyses are derived similarly to the $14\tev$ case, using 
exactly the same cuts as in the $14\tev$ case.

\begin{table}
\begin{center}
\begin{tabular}{|c|c|c|c|c|c|c|c|c|c|c|}
\cline{1-11}
\multirow{2}{*}{$\sigma$ (fb)}& \multicolumn{2}{|c|}{QCD $Zjj$} & \multicolumn{2}{|c|}{QCD $Wjj$} & \multicolumn{2}{|c|}{EW $Zjj$} & \multicolumn{2}{|c|}{EW $Wjj$} & \multicolumn{2}{|c|}{Total} \\ \cline{2-11}
&\cite{Eboli:2000ze}& Here &\cite{Eboli:2000ze}& Here
&\cite{Eboli:2000ze}& Here &\cite{Eboli:2000ze}& Here
&\cite{Eboli:2000ze}& Here \\ \cline{1-11}
Eqs.\ (\ref{Jetcuts}-\ref{Mcuts}) & 1254 & 1055 & 1284 & 906 & 151 & 148 & 101 & 85 & 2790 & 2194 \\ \cline{1-11}
Eqs.\ (\ref{Jetcuts}-\ref{phicuts}) + C.J.V. & 71.8 & 56.6 & 70.2 & 47.3 & 14.8 & 14.6 & 9.9 & 8.2 & 167 & 127  \\ \cline{1-11}
\hline

\end{tabular}
\caption{Total cross section (in fb) for various background components
  after applying particular sets of cuts. The results found in the
  work \cite{Eboli:2000ze} are presented along with those found in this
  work for comparison. Here ``C.J.V.'' refers to application of survival probabilities
  for the central jet veto, as described in \cite{Chehime:1992ub}.}
\label{tableBG}
\end{center}
\end{table}

\subsection{Selection for WBF at LHC: Signal Subprocess Cross-Section}
\label{SignalWBF}
\begin{figure}[hbtp]
    \centering
    \includegraphics[width=1.0\textwidth]{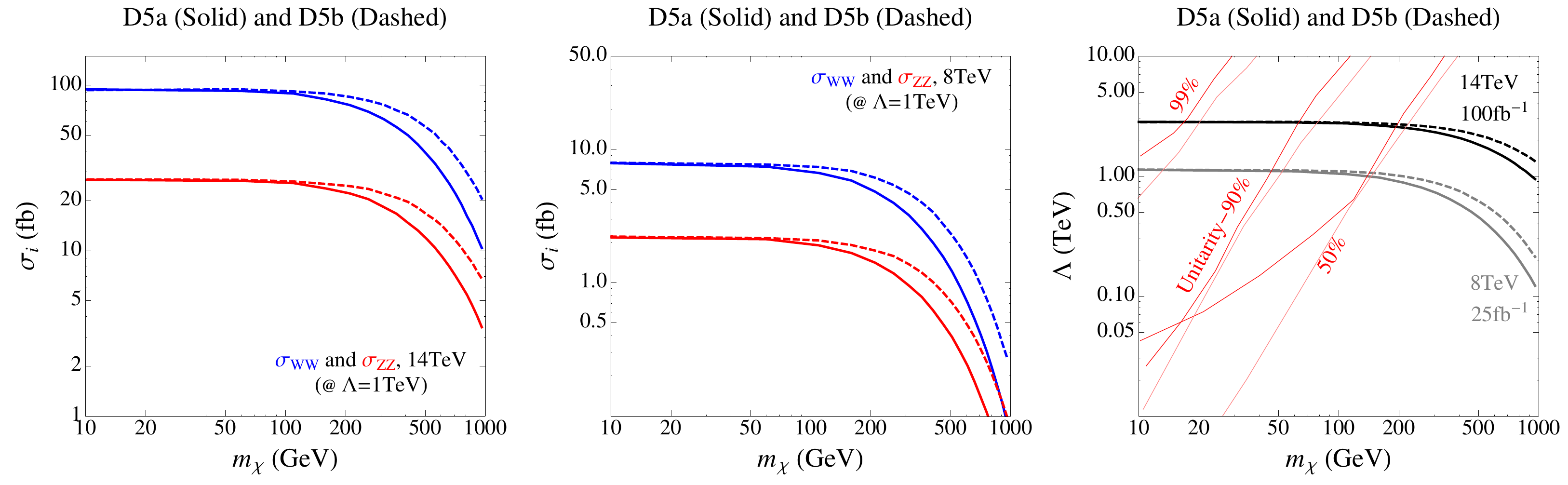}
    \caption{We display subprocess cross-sections (with $\Lambda=1\tev$) for the $d=5$ ``Higgs-portal''
      operators D5a (solid) and D5b (dashed). $W^+W^-$ and $Z^0Z^0$ subprocesses are described by blue
      and red curves, respectively. Results for $14\tev$ and $8\tev$ scenarios are shown in the left and 
      center panels, respectively. In the right panel we display
      curves representing 95\% CL lower limits on $\Lambda$ due to the subprocess combinations described for these
      operators in Eqn.\ \ref{combinations}. Bounds are set assuming either $14\tev$ $100\;\rm{fb}^{-1}$
      (black) or $8\tev$ $25\;\rm{fb}^{-1}$ (grey) WBF analyses. In both cases limits
      are computed assuming 5\% systematic uncertainty on the background. Red and pink curves describing the unitarity
      of the EFT description are also included, as explained in the text.}
    \label{figs:D5ab}
  \end{figure}

\begin{figure}[hbtp]
    \centering
    \includegraphics[width=1.0\textwidth]{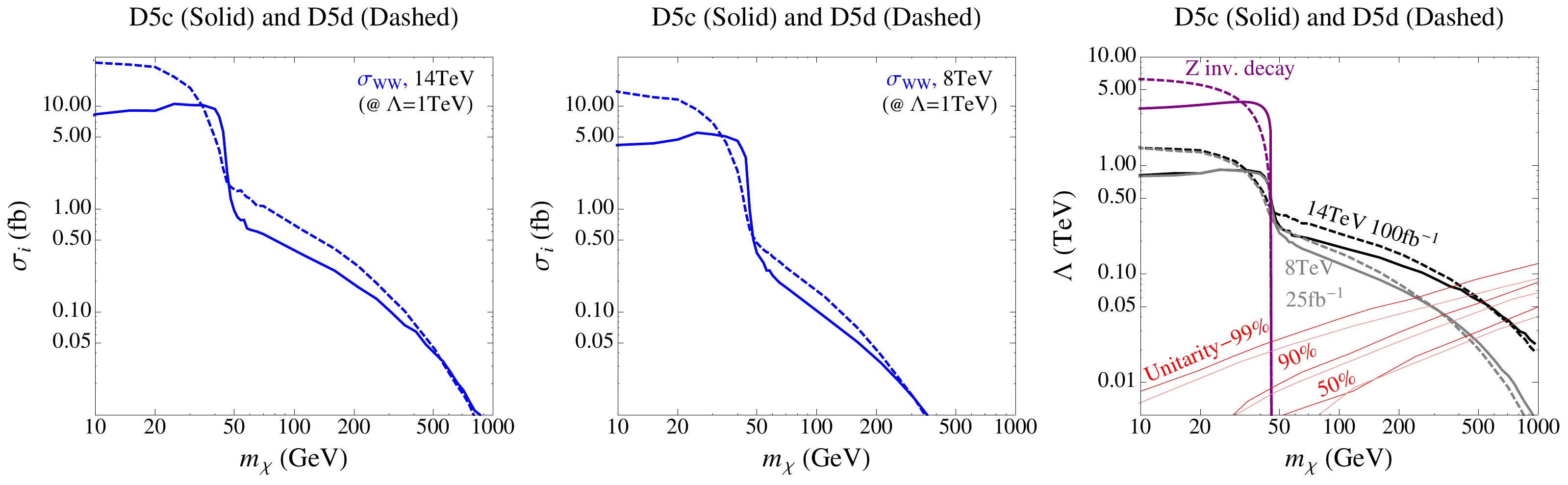}
    \caption{Similar to previous figure but now for the $d=5$ ``moment''
      operators D5c (solid) and D5d (dashed). For these operators the only relevant 
      subprocess is DM production through $W^+W^-$ and can occur via either 4-point
      or 3-point interactions. In the right panel we plot the limit from
      the $Z^0$ invisible decay width (purple) along with the results for
      $14\tev$ $100\;\rm{fb}^{-1}$ (black) or $8\tev$ $25\;\rm{fb}^{-1}$ (grey) WBF
      analyses. Red and pink curves describing the unitarity
      of the EFT description are also included, as explained in the text.}
    \label{figs:D5cd}
  \end{figure}

\begin{figure}[hbtp]
    \centering
    \includegraphics[width=1.0\textwidth]{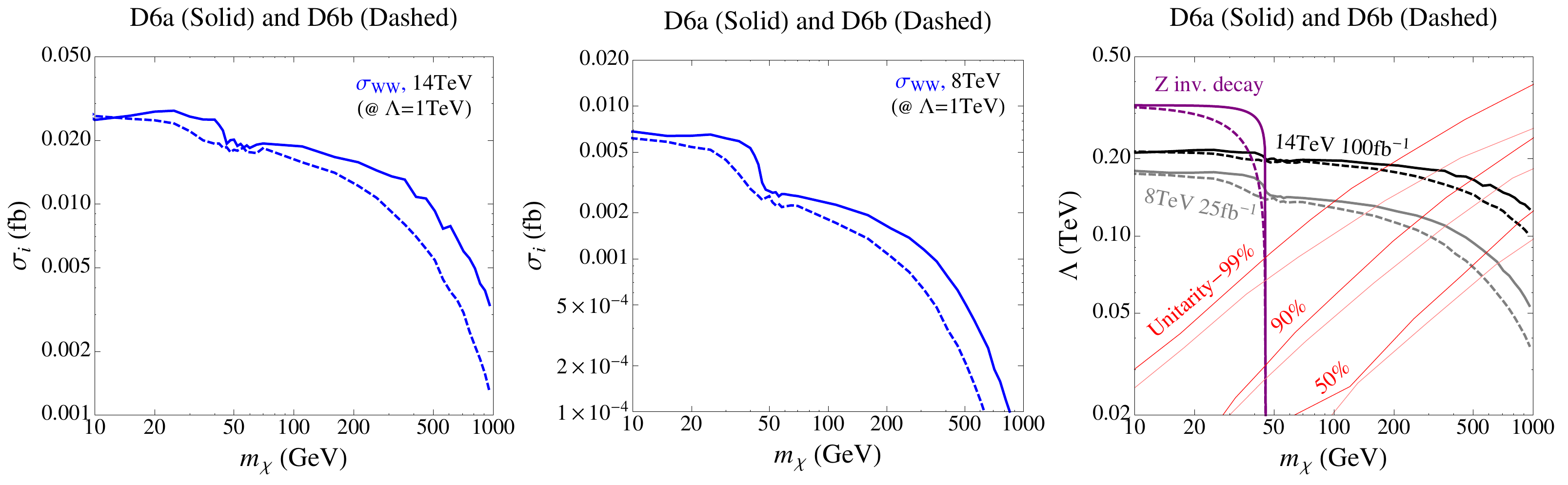}
    \caption{Same as in the previous figure but now for the $d=6$
    operators D6a (solid) and D6b (dashed).}
    \label{figs:D6ab}
  \end{figure}

\begin{figure}[hbtp]
    \centering
    \includegraphics[width=1.0\textwidth]{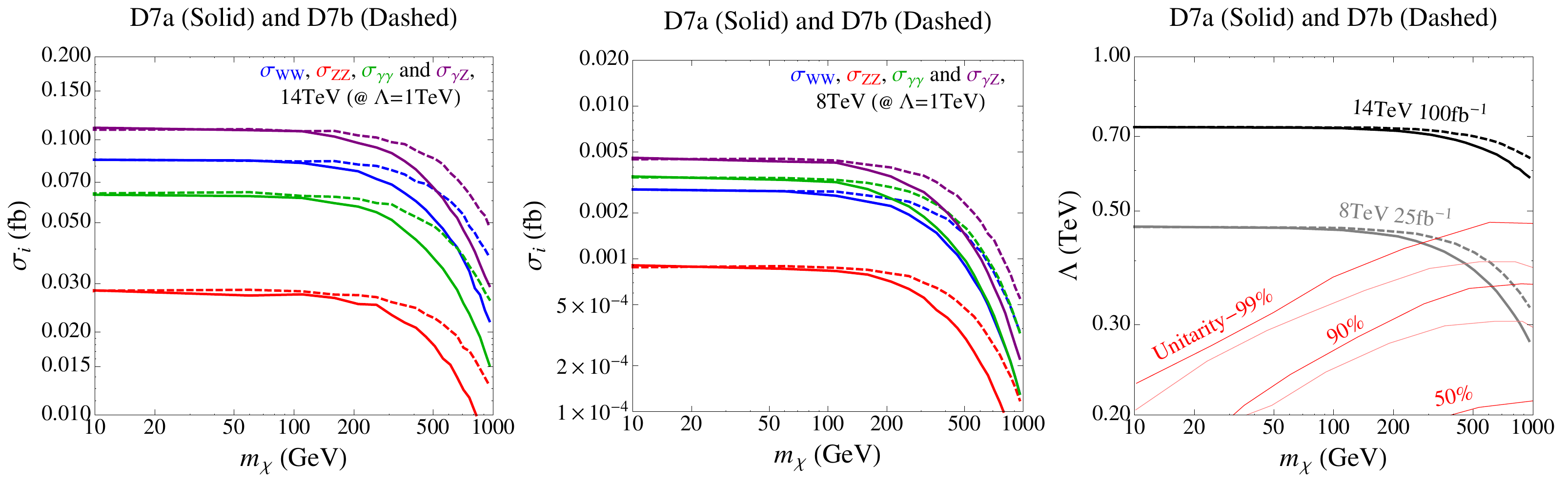}
    \caption{We display subprocess cross-sections (with
    $\Lambda=1\tev$) for the $d=7$ operators D7a (solid) and D7b
    (dashed). The subprocesses $W^+W^-$, $Z^0Z^0$, $\gamma\gamma$
    and $\gamma Z^0$ may all result from this class of operators and
    are represented here by blue, red, green and purple curves, respectively. 
    Results for $14\tev$ and $8\tev$ scenarios are shown in the left and
    center panels, respectively. In the right panel we display
    curves representing 95\% CL lower limits on $\Lambda$ due to the subprocess 
    combinations described for these operators in Eqn.\ \ref{combinations}. Bounds
    are set assuming either $14\tev$ $100\;\rm{fb}^{-1}$
    (black) or $8\tev$ $25\;\rm{fb}^{-1}$ (grey) WBF analyses. In both cases limits
    are computed assuming 5\% systematic uncertainty on the background. Red and pink
    curves describing the unitarity of the EFT description are also included, as explained in the text.}
    \label{figs:D7ab}
  \end{figure}

\begin{figure}[hbtp]
    \centering
    \includegraphics[width=1.0\textwidth]{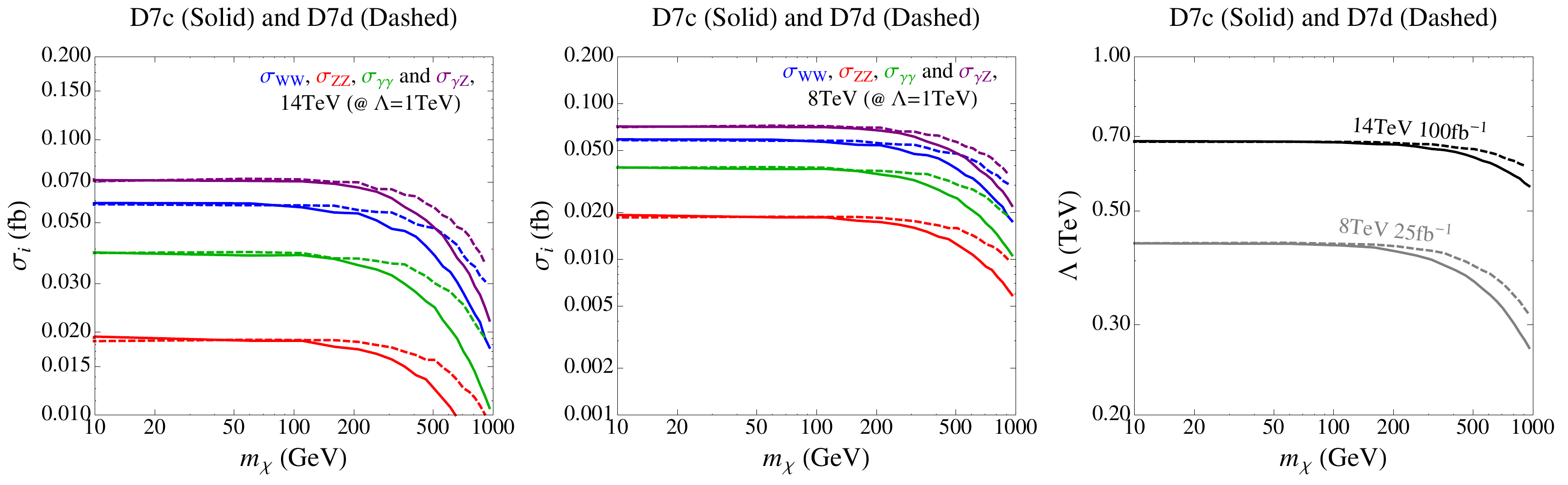}
    \caption{Similar to the previous figure but now for operators D7c (solid) and D7d (dashed). Unitarity
    curves were not calculated for these operators but are expected to be similar to those found for the operators D7a,b.}
    \label{figs:D7cd}
  \end{figure}

 Signal cross-sections are generated using the same procedure
as that described in Section \ref{BGWBF} for generating SM
backgrounds. We use FeynRules \cite{Christensen:2008py}\footnote{FeynRules is a Mathematica package
  that allows for the calculation of momentum space Feynman rules for
  quite generic models of new physics (specified at the Lagrangian
  level). Here we have been able to interface FeynRules and Madgraph v5
 using the Universal FeynRules Output (UFO) language.} in concert with
Madgraph v5 \cite{Alwall:2011uj} to simulate signal events from our
effective contact interactions. We have generated all signal subprocess
cross-sections with $\Lambda=1\tev$. The signal cross-sections are 
$\propto 1/\Lambda_s^{2n}$ with n=1, 2 and 3 for contact operators of
dimension $d=5$, $6$ and $7$, respectively, so we can obtain results for
$\Lambda_s\neq1\tev$ by the appropriate scaling. Subprocess
cross-sections (for $\Lambda=1\tev$) and 95\% CL lower limits on $\Lambda$
are displayed in Figures \ref{figs:D5ab}-\ref{figs:D7cd}. 

In cases where
an operator is associated with multiple subprocesses we have to choose 
a definite weighting for the subprocesses in order to plot limits on
$\Lambda$. The combinations that were chosen to produce 
Figs.\ \ref{figs:D5ab}-\ref{figs:D7cd} are:
\begin{align}
\label{combinations}
&\mathrm{D5a:}&\qquad \frac{1}{\Lambda}\bar{\chi}\chi\left(\frac{Z^{\mu}Z_{\mu}}{2} + W^{+\mu}W^{-}_{\mu} + h.c.\right) \\\nonumber
&\mathrm{D5b:}&\qquad \frac{1}{\Lambda}\bar{\chi}i\gamma_5\chi\left(\frac{Z^{\mu}Z_{\mu}}{2} + W^{+\mu}W^{-}_{\mu} + h.c.\right) \\\nonumber
&\mathrm{D5c:}&\qquad \frac{g_w}{\Lambda}\left(\bar{\chi}\sigma_{\mu\nu}t^3\chi W^{3\mu\nu}+\frac{s_w}{c_w}\frac{Y}{2}\bar{\chi}\sigma_{\mu\nu}\chi B^{\mu\nu}\right) \\\nonumber
&\mathrm{D5d:}&\qquad \frac{g_w}{\Lambda}\left(\bar{\chi}\sigma_{\mu\nu}t^3\chi \widetilde{W}^{3\mu\nu}+\frac{s_w}{c_w}\frac{Y}{2}\bar{\chi}\sigma_{\mu\nu}\chi \widetilde{B}^{\mu\nu}\right)\\\nonumber
&\mathrm{D6a:}&\qquad \frac{g_w}{\Lambda^2}\left(\bar{\chi}\gamma_{\mu}t^3 D_{\nu}\chi W^{3\mu\nu} + \frac{s_w}{c_w}\frac{Y}{2}\bar{\chi}\gamma_{\mu} D_{\nu}\chi B^{\mu\nu} \right)\\\nonumber
&\mathrm{D6b:}&\qquad \frac{g_w}{\Lambda^2}\left(\bar{\chi}\gamma_5\gamma_{\mu}t^3 D_{\nu}\chi W^{3\mu\nu} + \frac{s_w}{c_w}\frac{Y}{2}\bar{\chi}\gamma_5\gamma_{\mu} D_{\nu}\chi B^{\mu\nu} \right)\\\nonumber
&\mathrm{D7a:}&\qquad \frac{1}{\Lambda^3}\bar{\chi}\chi W^{a\mu\nu}W^a_{\mu\nu} \\\nonumber
&\mathrm{D7b:}&\qquad \frac{1}{\Lambda^3}\bar{\chi}i\gamma_5\chi W^{a\mu\nu}W^a_{\mu\nu} \\\nonumber
&\mathrm{D7c:}&\qquad \frac{1}{\Lambda^3}\bar{\chi}\chi W^{a\mu\nu}\widetilde{W}^a_{\mu\nu} \\\nonumber
&\mathrm{D7d:}&\qquad \frac{1}{\Lambda^3}\bar{\chi}i\gamma_5\chi W^{a\mu\nu}\widetilde{W}^a_{\mu\nu}.
\end{align}
The combinations shown for operators D5c-d and D6a-b are fixed by the requirement that the $\chi\chi A^0$
vertex vanishes but all of the other expressions have been chosen arbitrarily for simplicity. In general
one can weight the cross-section curves given in Figs.\ \ref{figs:D5ab}-\ref{figs:D7cd} and sum to get a 
total cross-section according to the details of their own UV complete theory as described in Section 
\ref{sec:method}.

The general features of the curves in Figs.\ \ref{figs:D5ab}-\ref{figs:D7cd} are easy to understand: sensitivity is constant for all DM masses 
$m_{\chi}\ll\sqrt{\hat{s}}$ and falls dramatically at $m_{\chi}\sim 1\tev\sim\sqrt{\hat{s}}$. There is
little difference in pairs of operators that differ only by a $\gamma_5$, except where the phase space 
of the DM particles becomes important near $m_{\chi}\sim\sqrt{\hat{s}}$. The differences in overall reach
between the various operators arise not only from the basic fact that they have different naive scaling dimensions, 
but also because of kinematic differences that have an impact on the effectiveness of our WBF cuts. The operators
 D5c-d and D6a-b are all operators for which we have both 3-point and 
4-point contact interactions. The 3-point vertices are problematic here, as they result in diagrams
which are kinematically similar to the SM backgrounds (Fig.\ \ref{figs:BGfeyn}). In these cases we
expect our WBF search to perform significantly worse as our cuts are not as able to distinguish 
signal and background. In Figs.\ \ref{figs:D5ab}-\ref{figs:D7cd} we also show the regions 
constrained\footnote{Since the requirement that our DM is not milli-charged fixes the relationship
 between the coefficients of $SU(2)_L$ and hypercharge gauge bosons in our EFT description
 one has no way of getting around this invisible width constraint unless somehow the very
 tight constraints \cite{McDermott:2010pa} on milli-charged DM can be avoided.} by the measurement
 of the invisible width of the $Z^0$.

\begin{figure}[hbtp]
    \centering
    \includegraphics[width=0.80\textwidth]{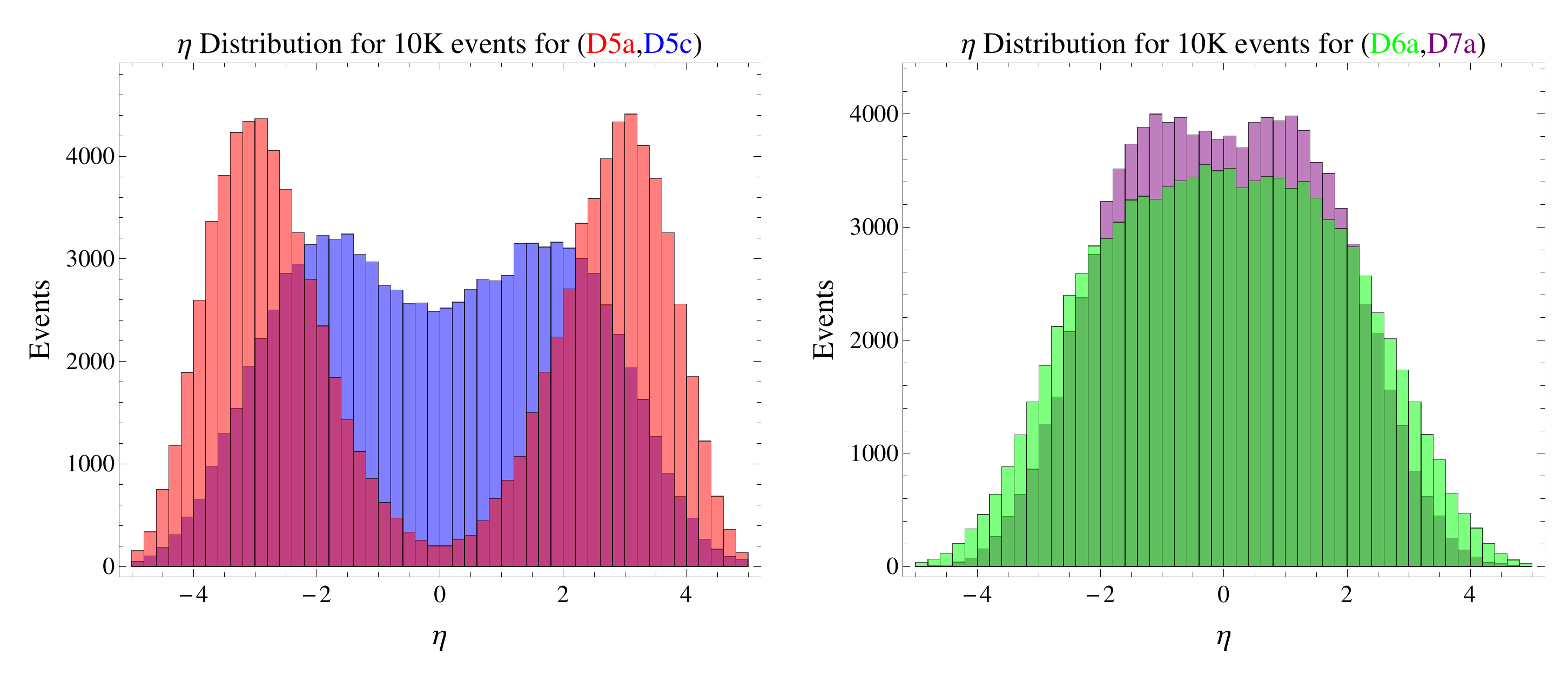}
    \caption{We display the jet rapidity distributions of various operators for 10,000 events, 
      \ie~ they are not normalized to their respective cross-sections.}
    \label{figs:etaSN}
  \end{figure}

Although our $d=7$ operators do not
suffer from the above issue, their WBF signal events are also somewhat difficult to separate from the backgrounds.
This happens because the D7 operators are \emph{not} predominantly produced from longitudinally polarized $W$
bosons, which are both the dominant source of electroweak gauge bosons in the beam protons and are
preferentially supplied with low transverse momentum. Thus the signal event rates are somewhat lower and the
leading jets from WBF production of the D7a-d operators tend to be more central, suppressing signal efficiencies 
for the leading jet cuts. For illustration leading jet rapidity distributions are shown for many of our operators
in Figure \ref{figs:etaSN}.

In Figures \ref{figs:D5ab}-\ref{figs:D7cd} we also display curves describing the unitarity of our effective operator description. 
We include curves for which approximately $99\%$, $90\%$, or $50\%$ of simulated events appear \emph{not} to violate unitarity (red and
pink curves are appropriate for $14\tev$ and $8\tev$ events, respectively). Roughly these curves separate regions \emph{above} which our
EFT is an excellent ($99\%$), good ($90\%$) or poor ($50\%$) description as far as unitarity is concerned. These curves were calculated
semi-analytically, combining the closed-form longitudinally polarized amplitudes for each operator (as calculated using FeynArts 3.4 
\cite{Hahn:2000kx}) with the approximate analytical parton luminosity for longitudinally-polarized W-bosons derived in 
\cite{Cahn:1983ip} and CTEQ5M parton distribution functions. Although the unitarity curves for operators 
related simply by a $\gamma_5$ are somewhat different we show only the curves associated to the ``non-$\gamma_5$'' operators in Figures \ref{figs:D5ab}-\ref{figs:D7ab}
in order to reduce clutter (curves for the operators with $\gamma_5$'s are displayed later on in Figs.\ \ref{figs:comboD5ab}-\ref{figs:comboD7ab}).
We see that the Higgs Portal operators D5a,b are the most challenged in this regard, with the WBF excluded regions
reaching above the ``good'' or ``excellent'' curves only for very light $m_{\chi}\lsim 50\gev$. The dipole moment and vector exchange
operators D5c,d appear to be good descriptions up to $m_{\chi}$ of several hundred $\gev$ and the $d=7$ operators have no
problems with unitarity up until the kinematic reach of the searches $m_{\chi}\sim 1\tev$.

\begin{figure}[hbtp]
  \centering
  \includegraphics[width=1.0\textwidth]{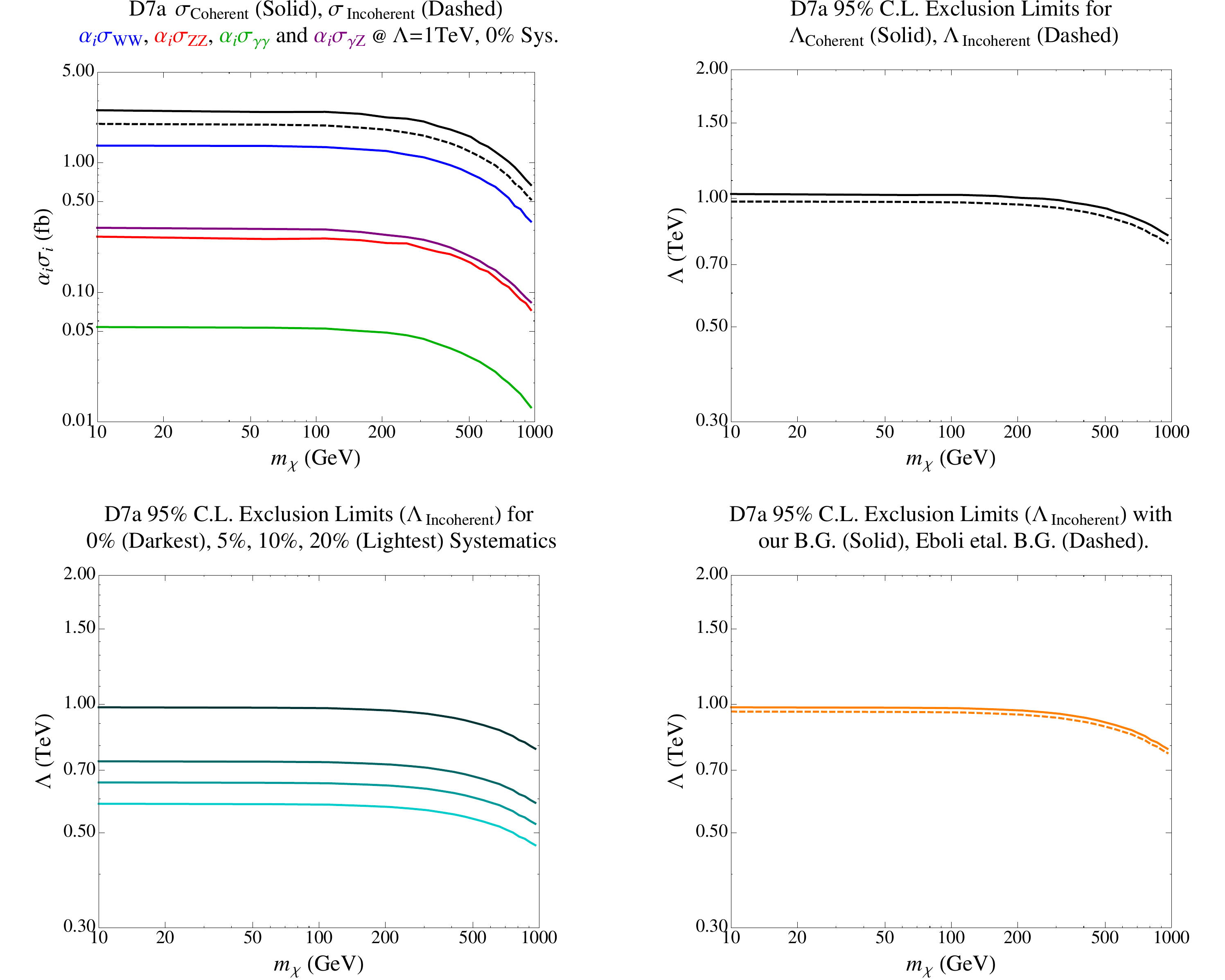}
  \caption{We display WBF bounds for $14\tev$/100 $\rm{fb}^{-1}$
LHC analyses on the operator D7a. In the upper-left panel we
display weighted subprocess cross-sections (as described in the
figure) along with the coherent (black-solid) and incoherent
(black-dashed) total signal cross-sections. In the upper-right panel
we translate coherent and incoherent total cross-sections into bounds
on the high scale $\Lambda$. In the lower-left panel we describe the
effect of background systematic uncertainty on the $\Lambda$ bound,
showing 0\%, 5\%, 10\% and 20\% uncertainty, as denoted in the figure.
  In the lower-right panel we show the difference in bounds that can
  be placed using the backgrounds derived in this analysis (orange-solid) or in
  using the backgrounds derived in the work \cite{Eboli:2000ze}
  (orange-dashed).}
  \label{figs:example}
\end{figure}

Now let us follow up on our earlier discussion (Section \ref{sec:method}) of errors
and uncertainties in our WBF analysis, employing as an example the operator D7a. We would like
to illustrate the effect that the various sources of error/uncertainty
that we have been discussing, $\eg$, the incoherent sum of subprocesses,
systematic uncertainties on the background and differences between the
backgrounds as calculated here and elsewhere, in the context of this example.
 The panels of Figure \ref{figs:example} address each of these effects in turn.

In the upper-left panel of Fig.\ \ref{figs:example} we display
subprocess cross-sections with weighting ($eg$, $\alpha_i \sigma_i$, with $\alpha_i$
as in Eq.\ (\ref{D7sigmatot})), taking for simplicity $a=1$ and $b=0$. We
also display the total cross sections resulting from coherent and
incoherent summation of the subprocesses, $\sigma_{coherent}$ and
 $\sigma_{incoherent}$. In the upper-right panel we convert these
total cross-sections into 95\% C.L. lower limits on $\Lambda$,
$\Lambda_{coherent}$ and $\Lambda_{incoherent}$. We observe that there
 is constructive interference between the different subprocesses in
 this example, resulting in a fractional error of $\approx 20\%$ in
 using the approximation $\sigma_{tot}\approx\sigma_{incoherent}$.
We note, however, that the fractional error induced in using the
bound $\Lambda_{incoherent}$ instead of $\Lambda_{coherent}$ is
related to the fractional error in the total cross-section as
\begin{eqnarray}
\frac{\delta\Lambda}{\Lambda} =
\frac{1}{2n}\frac{\delta\sigma}{\sigma},
\label{Lambdafrac}
\end{eqnarray}
(again, n=1, 2 and 3 for contact operators of dimension $d=5$, 6 and 7,
 respectively) so that, $\eg$, a $20\%$ error due to neglecting 
interference in $\sigma_{tot}$ will induce an error
in $\Lambda$ of only $5\%$. This is reflected in the upper-right panel of figure
\ref{figs:example}.

In the lower-left panel of Fig.\ \ref{figs:example} we display
95\% C.L. lower limits, in all cases using
$\Lambda_{incoherent}$, for four different values of assumed
systematic uncertainty on the background (0\%, 5\%, 10\% and
20\%). The variation in these curves is significantly larger than that
seen in the other panels.

In the lower-right panel of Fig.\ \ref{figs:example} we display
95\% C.L. lower limits (in all cases using
$\Lambda_{incoherent}$ and ignoring the systematic uncertainty) employing either the
background rates found in this analysis or the background rates found in the
work \cite{Eboli:2000ze}. As mentioned previously, for a given assumed systematic uncertainty, the difference in bounds
resulting from using the backgrounds derived in \cite{Eboli:2000ze} and in
using the backgrounds derived here is relatively insignificant. Overall we see that systematic uncertainty is expected 
to be the dominant source of error in our analysis.

\section{Dark Matter Search Bounds}
\label{sec:dm}
 Here we investigate bounds on our contact operators that can be derived from astrophysical data.
As we discussed in Section \ref{sec:operators}, signals
in experiments which probe our contact interaction via non-relativistic DM scattering or annihilation are particularly well-modeled by
an effective operator description, in possible contrast to WBF bounds derived in this context. Of course, DM signals from astrophysical
DM distributions are also subject to many sources of uncertainty, $\eg$, in estimating average DM relic abundance \cite{Gates:1995dw},
local effects of DM substructure \cite{Diemand:2008in}-\cite{Delahaye:2010zz}, propagation of SM products of DM annihilation, etc.,
 that are not present in collider searches. As we discussed earlier, we do not expect significant direct detection scattering rates from
our operators as operators with only 4-point contact interactions must scatter through higher order processes and as operators that generate
the $\chi\chi Z$ three-point vertex are momentum suppressed. Given this, our primary focus here will be to derive bounds that can be set from null
 searches for DM annihilation in our Milky Way (MW) DM halo.

Earth-bound and satellite-born detectors search for the products of DM annihilations in the MW halo by measuring a variety
of energetic particle spectra. Our operators produce a wide variety of SM final states ($W^+W^-$, $Z^0Z^0$, $\gamma Z^0$ and
$\gamma\gamma$ for operators with 4-point interactions only and additional fermionic final states for 3-point interactions where
DM annihilates through s-channel $\gamma/Z^0$ ) and thus can be constrained by a variety of indirect detection experiments. Here
we focus on three classes of search: $\gamma$-ray spectral limits from MW dwarf spheroidal galaxies, monochromatic $\gamma$-ray
line searches in the MW halo, and measurements of antiproton cosmic-ray spectra. 

\subsection{Indirect Detection Bounds}
Dwarf spheroidal galaxies are extremely DM dominated
 satellites of the MW which, having essentially zero intrinsic astrophysical $\gamma$-ray sources, are an excellent place to look for
the continuum $\gamma$-ray spectra that accompanies DM annihilation into essentially all final states. For this bound we employ the
95\% C.L. limits set by the \textit{Fermi}-LAT collaboration using a combination of observations of ten MW dwarf-spheroidals \cite{Ackermann:2011wa}, as well as 
limits derived from the VERITAS collaboration's observations of the MW dwarf Segue-I \cite{Aliu:2012ga}. The low-energy threshold of the 
satellite-born LAT instrument extends far below the $\sim 100 \gev$ threshold of the earth-bound VERITAS air Cerenkov telescope array
so that the LAT sets the most stringent limits for DM masses below $\sim 900\gev$. Above this, the large fiducial volume of VERITAS
(the atmosphere above the array) results in the tightest constraints on heavier DM. As the results quoted in \cite{Ackermann:2011wa}\cite{Aliu:2012ga}
 are described in terms of limits on 
cross-sections into specific SM final-state channels ($WW$, $b\bar{b}$, $\tau\bar{\tau}$ and $\mu\bar{\mu}$) we do not have to model the signal $\gamma$-ray 
spectra or the DM distributions in the dwarfs here. As has been observed \cite{Cotta:2011pm}, the continuum $\gamma$-ray spectra from annihilations to $ZZ$ lead
 to essentially the same limits as
that from annihilation to $WW$ and that annihilations to all light quarks produce essentially the same limits as that from annihilation to $b\bar{b}$. 
Given this, we sum the cross-sections for annihilation to the $WW$ and $ZZ$ final states and those for all light quarks in comparing to the experimental
 $WW$ and $b\bar{b}$ limit curves, respectively. In order to calculate the relic density, total annihilation cross-section and cross-sections for 
particular final state channels we use FeynRules 1.6.0 to calculate the Feynman rules for each operator and interface with MicrOMEGAs 2.2 \cite{Belanger:2008sj}
to calculate the DM observables.

We use limits on $\gamma$-ray lines that were set by the \textit{Fermi}-LAT collaboration \cite{Ackermann:2012qk}, to bound operators
that can annihilate directly into the $\gamma\gamma$ and $\gamma Z^0$ and final states. We assume an NFW profile \cite{NFW} for the MW DM halo. Results
for the more conservative isothermal profile would be about 30-40\% less constraining. Relating the DM mass to the $\gamma$-ray line energy,
the LAT data provide constraints on $7\gev<m_{DM}<200\gev$ DM annihilating to $\gamma\gamma$ and on $60\gev<m_{DM}<210\gev$ DM annihilating to
$\gamma Z^0$.

 We use the PAMELA collaboration's measurement \cite{Adriani:2010rc} of the ratio of cosmic-ray antiprotons to protons to bound annihilations
 producing substantial hadronic matter. We again assume an NFW profile for the DM halo in calculating our antiproton signal rates and setting
 bounds. We use a modified\footnote{modified to use non-SUSY models.}
 version of the numerical package DarkSUSY 5.0.4 to calculate the signal antiproton injection spectra. This injection spectra is then propagated to
 obtain a local signal flux spectrum using a propagation model (galdef\_50p\_599278) that is supplied in the GALPROP v50.1p package
 \cite{Strong:1998pw}\cite{Moskalenko:2001ya}\cite{GALPROP} and is seen
 to be a good fit to a variety of astrophysical observations. We calculate the bounds from the PAMELA data\footnote{Taking the 17 highest energy bins, as the
 lower energy bins are affected by solar modulation.} by calculating a $\chi^2$, where we only include \emph{signal} protons/antiprotons and dividing by the
 (well-measured) primary proton cosmic-ray spectrum, 
excluding regions at 95\% confidence ($\chi^2/16 \geq 1.724$). As we are not adding any astrophysically produced ``secondary antiprotons'' this
exclusion is somewhat conservative, though calculations done with background added are seen to provide similar bounds. It should be noted that the numerical
tables used in DarkSUSY to derive the injection spectrum were created by scanning DM masses in the range $10\gev - 5\tev$, so one has less 
confidence in the numerical accuracy when extrapolating beyond this range ($\eg$, in looking at light dark matter scenarios)\footnote{Of course,
one should also consider the range over which the underlying code that has generated these tables (PYTHIA V6 \cite{Sjostrand:2006za}) is accurate. The authors
of DarkSUSY note, in particular, that there is something like a factor of 2 uncertainty in the antiproton yields at low energies due to the 
lack of reliable low-energy antiproton data with which to tune PYTHIA \cite{Gondolo:2004sc}.}. 


In using astrophysical experiments to set bounds on our operators we must make some assumption about the relic abundance of our DM. In all of the figures
that follow we determine excluded regions by assuming that the dark matter relic density is $\Omega h^2_{\chi}=\Omega h^2_{\mathrm{WMAP}}\sim 0.114$ ($\ie$, that 
the $\chi$ particle makes up \emph{all} of DM). In the figures we include a (red-dashed) curve in the $\Lambda$ vs.\ $m_{\chi}$ plane where the annihilation
through the contact operators alone would yield a relic density, calculated assuming the usual thermal cosmological evolution, that matches the WMAP \cite{Komatsu:2010fb}
value. In any fully specified model, however, the true relic density of the DM may be greater or less than this thermal value. For example,
the relic density could be decreased by annihilating to dark sector states that later decay into the SM, or increased by non-thermal
cosmological evolution \cite{Kamionkowski:1990ni}\cite{Giudice:2000ex}\cite{Chung:1998ua} or co-annihilation with states that have
relatively inefficient annihilation and are not included in our EFT picture \cite{Profumo:2006bx}\cite{Servant:2002aq}\cite{Griest:1990kh}. The current
relic density of $\chi$ cannot be larger than the
WMAP value (or DM would overclose the universe), but may be smaller than the WMAP value, in which case limits coming from searches
for annihilating DM would be substantially less-constraining than what we show in our figures (annihilation signals scale as 
$(\Omega h^2_{\chi}/\Omega h^2_{\mathrm{WMAP}})^2$\;). If $\chi$ is imagined to be a Dirac fermion then the relative abundance of $\chi$ and $\bar{\chi}$ is 
important for determining limits. In this case we are assuming that the current relic abundance of $\chi$ is equal to the current relic abundance of $\bar{\chi}$. 
If this is not the case, and either the $\chi$ or $\bar{\chi}$ abundance dominates (as in asymmetric DM scenarios \cite{Kaplan:2009ag}-\cite{Fitzpatrick:2010em}),
then any bounds from annihilation would vanish. To zeroeth-order, the reach of indirect detection limits on a given operator just depends on whether
 the operator is velocity suppressed in the non-relativistic limit. Such operators are scaled by a factor $\upsilon^2\sim 10^{-6}$
compared to unsuppressed operators. A characterization of the leading order scaling of each operator can be found in the ``Annihilation'' column of
Table \ref{optable}. 

\subsubsection{Results}
\begin{figure}[hbtp]
    \centering
    \includegraphics[width=0.8\textwidth]{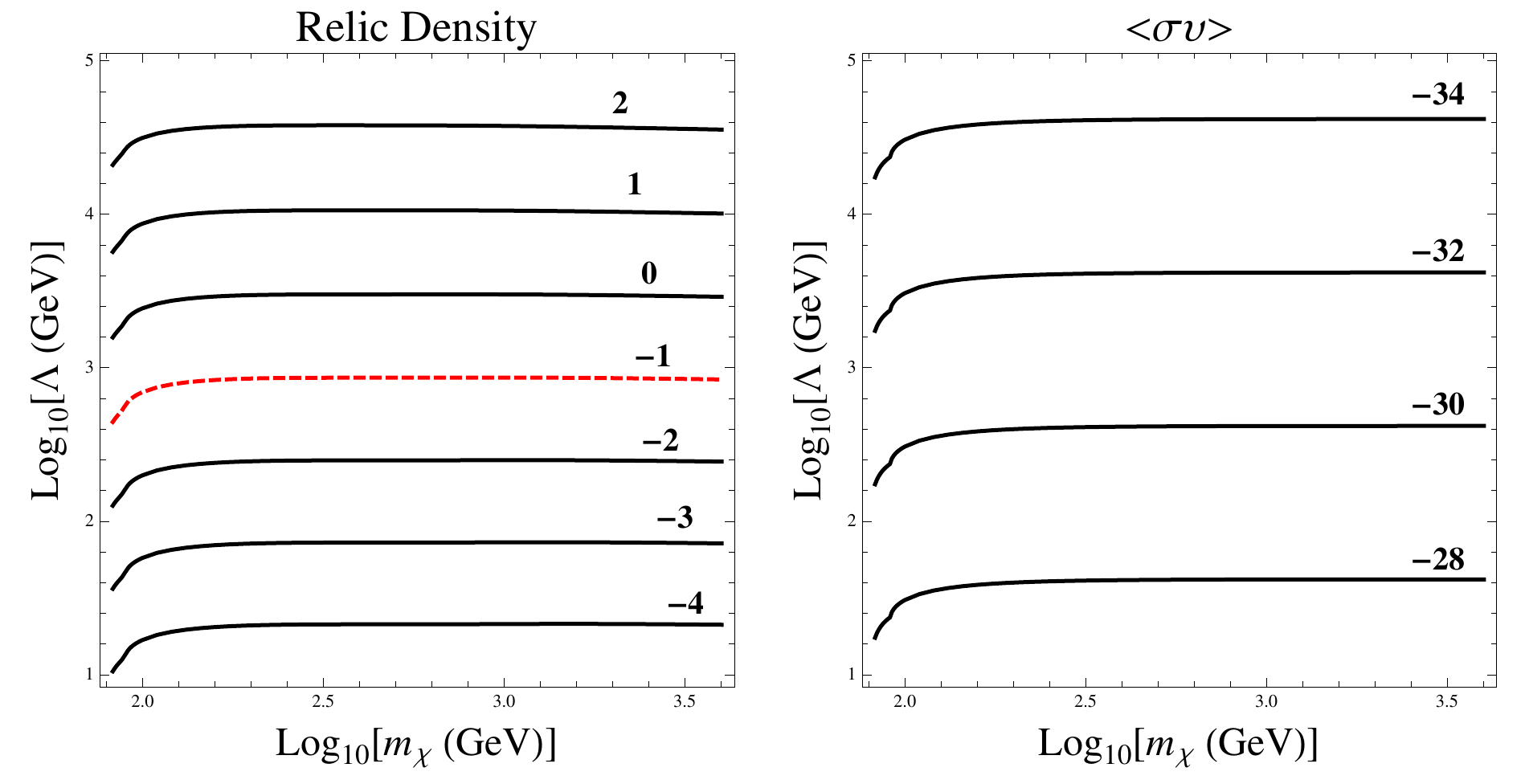}
    \caption{Relic density (left panel) and annihilation cross-section (right panel) contours  in the $\Lambda$ vs.\ $m_{\chi}$ plane for the Higgs-portal operator \textbf{D5a}.
      Contour labels are values of $\mathrm{log}_{10}(\Omega h^2_{\chi})$ and of $\mathrm{log}_{10}(\langle\sigma\upsilon\rangle~\mathrm{cm}^3\mathrm{s}^{-1})$,
      respectively. This operator is heavily velocity suppressed, so that, even under the assumption that $\Omega h^2_{\chi}\approx\Omega h^2_{WMAP}$, the
      current annihilation cross-section is much too low for any exclusion from indirect detection searches.
    }
    \label{figs:D5a}
  \end{figure}

\begin{figure}[hbtp]
    \centering
    \includegraphics[width=1.0\textwidth]{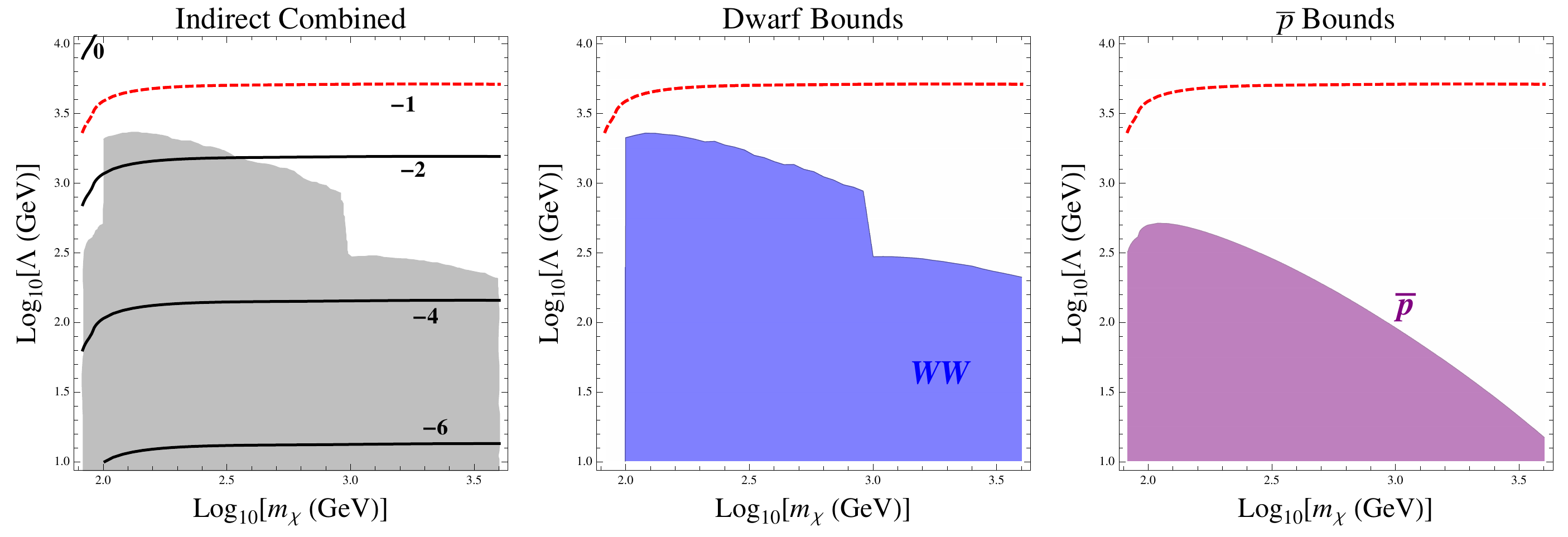}
    \caption{Indirect detection limits for the operator \textbf{D5b} in the $\Lambda$ vs.\ $m_{\chi}$ plane.
      The left panel includes contours of constant relic density (as calculated assuming thermal cosmology),
      labeled with values of $\mathrm{log}_{10}(\Omega h^2_{\chi})$, with the contour having $\Omega h^2_{\chi}\approx\Omega h^2_{WMAP}$
      represented by the red-dashed line. The region which is excluded by one or more indirect searches is shaded in grey. In the middle 
      and right panels we describe regions that are excluded by individual indirect limits.}
    \label{figs:D5b}
  \end{figure}

\begin{figure}[hbtp]
    \centering
    \includegraphics[width=1.0\textwidth]{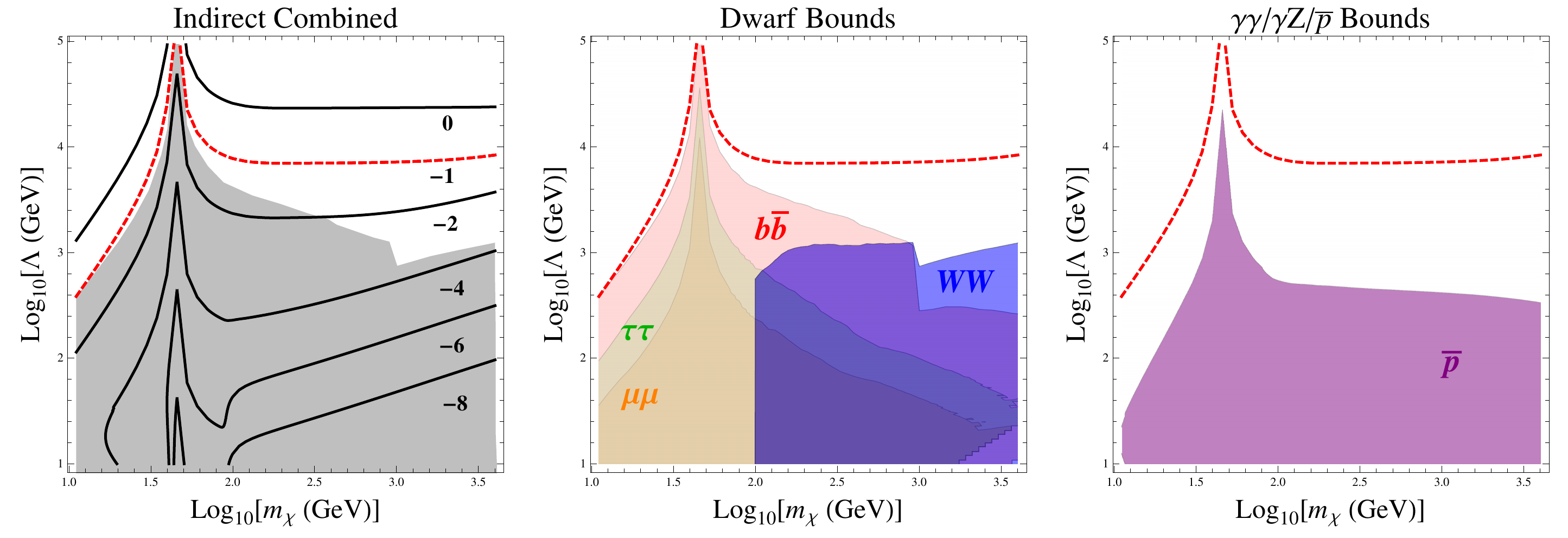}
    \caption{Same as in Figure \ref{figs:D5b} but for the operator \textbf{D5c}.}
    \label{figs:D5c}
  \end{figure}

\begin{figure}[hbtp]
    \centering
    \includegraphics[width=1.0\textwidth]{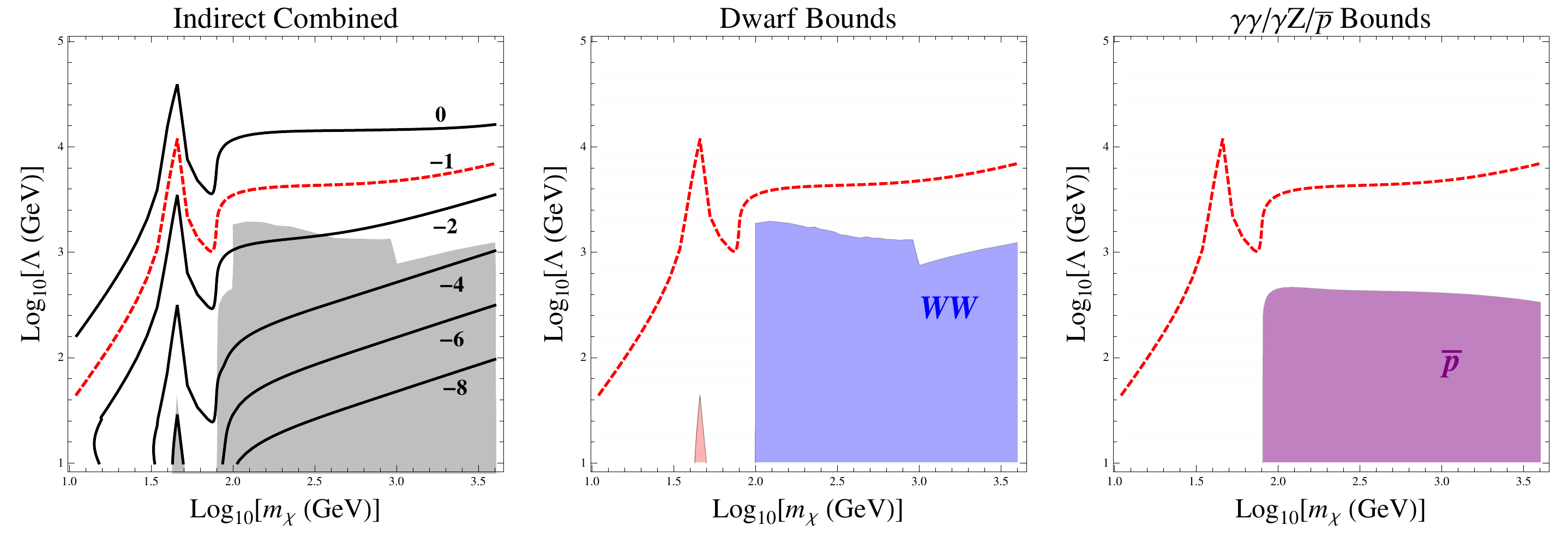}
    \caption{Same as in Figure \ref{figs:D5b} but for the operator \textbf{D5d}.}
    \label{figs:D5d}
  \end{figure}

\begin{figure}[hbtp]
    \centering
    \includegraphics[width=1.0\textwidth]{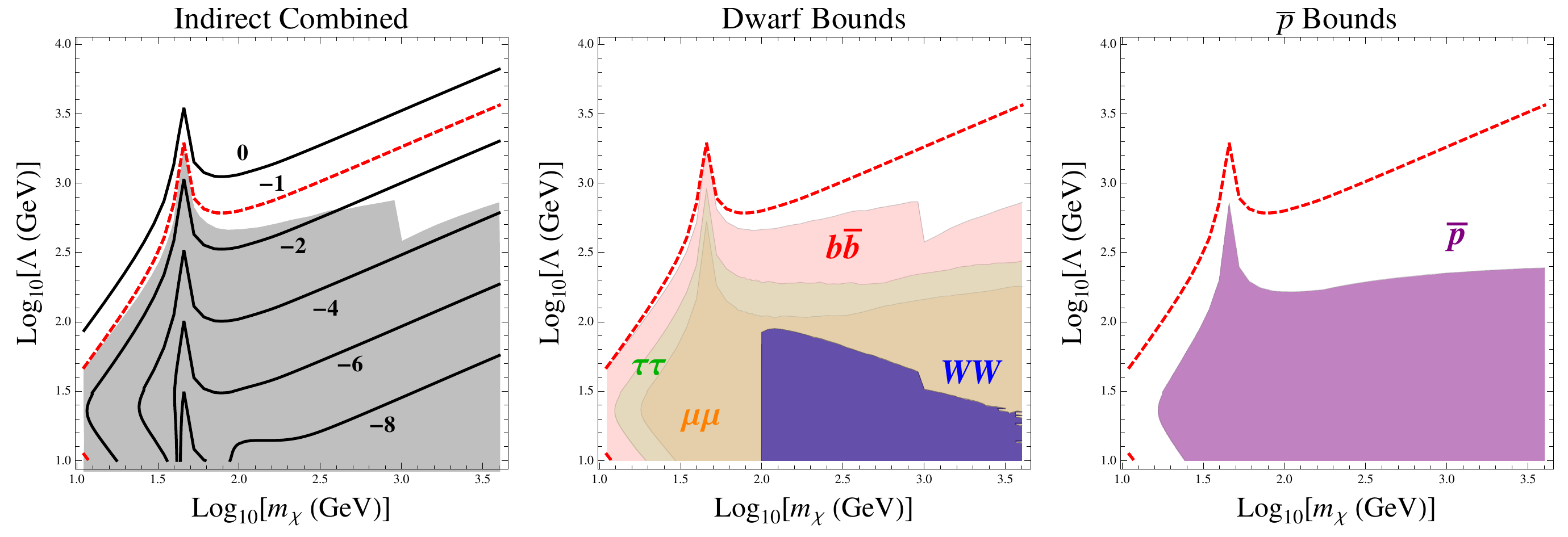}
    \caption{Same as in Figure \ref{figs:D5b} but for the operator \textbf{D6a}.}
    \label{figs:D6a}
  \end{figure}

\begin{figure}[hbtp]
    \centering
    \includegraphics[width=1.0\textwidth]{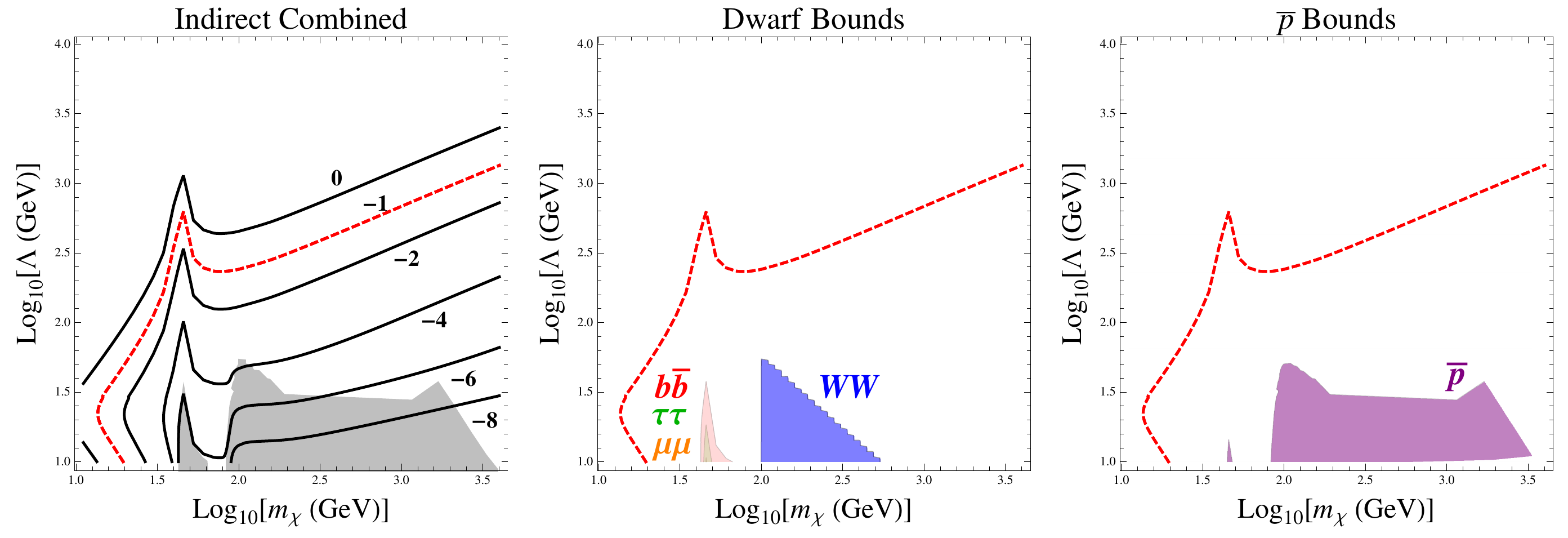}
    \caption{Same as in Figure \ref{figs:D5b} but for the operator \textbf{D6b}.}
    \label{figs:D6b}
  \end{figure}

\begin{figure}[hbtp]
    \centering
    \includegraphics[width=1.0\textwidth]{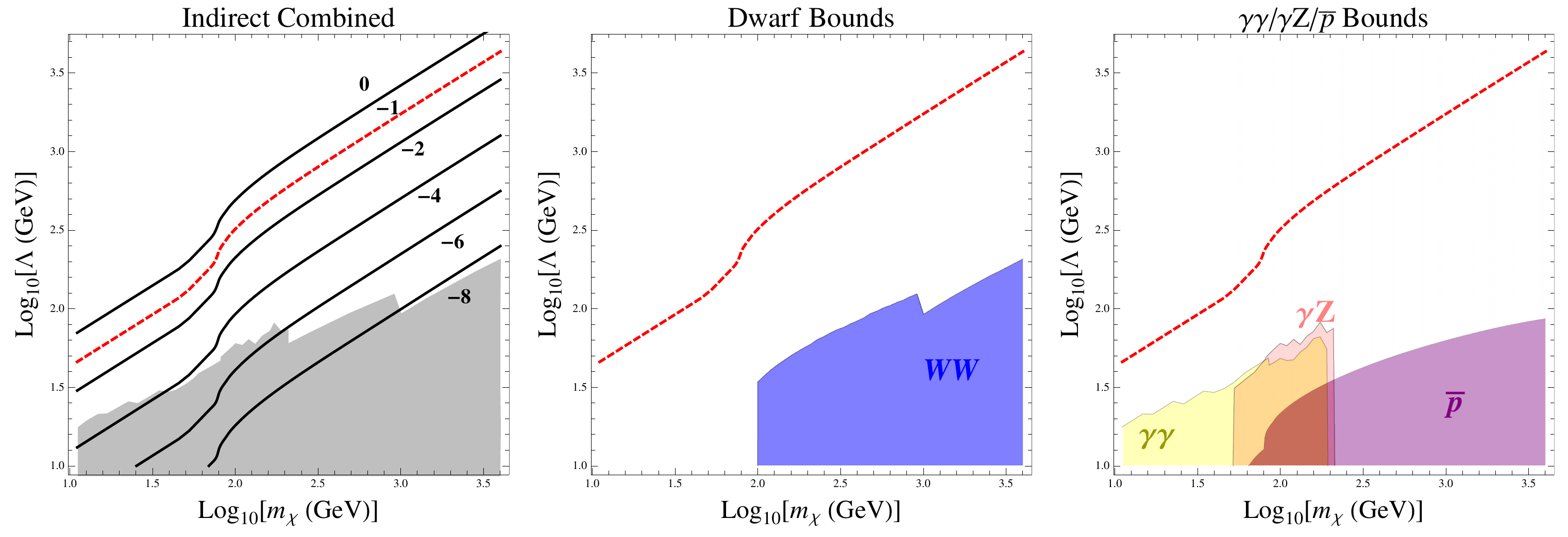}
    \caption{Same as in Figure \ref{figs:D5b} but for the operator \textbf{D7a}.}
    \label{figs:D7a}
  \end{figure}

\begin{figure}[hbtp]
    \centering
    \includegraphics[width=1.0\textwidth]{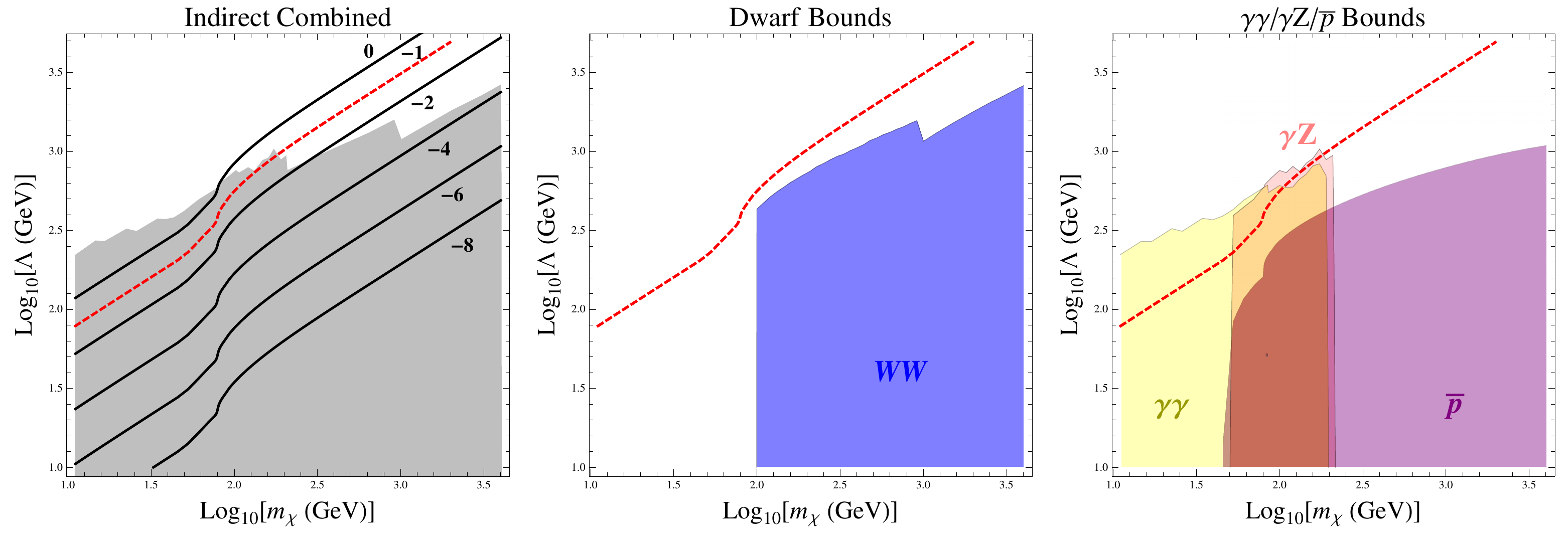}
    \caption{Same as in Figure \ref{figs:D5b} but for the operator \textbf{D7b}.}
    \label{figs:D7b}
  \end{figure}

\begin{figure}[hbtp]
    \centering
    \includegraphics[width=1.0\textwidth]{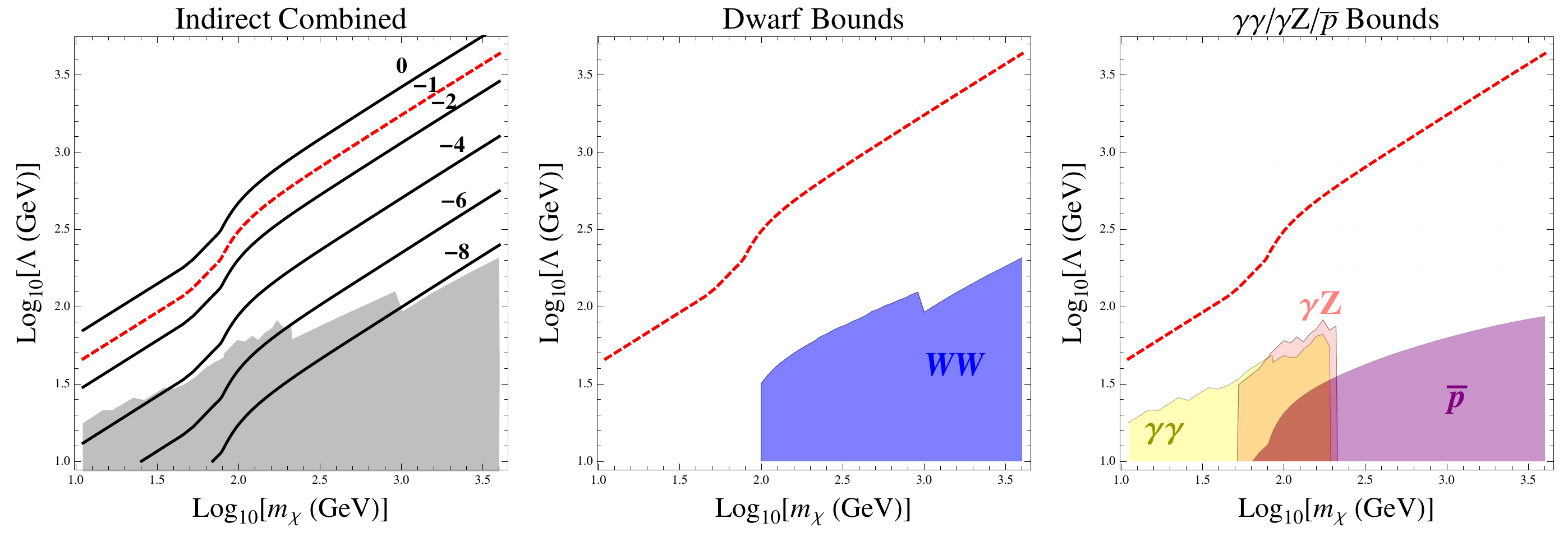}
    \caption{Same as in Figure \ref{figs:D5b} but for the operator \textbf{D7c}.}
    \label{figs:D7c}
  \end{figure}

\begin{figure}[hbtp]
    \centering
    \includegraphics[width=1.0\textwidth]{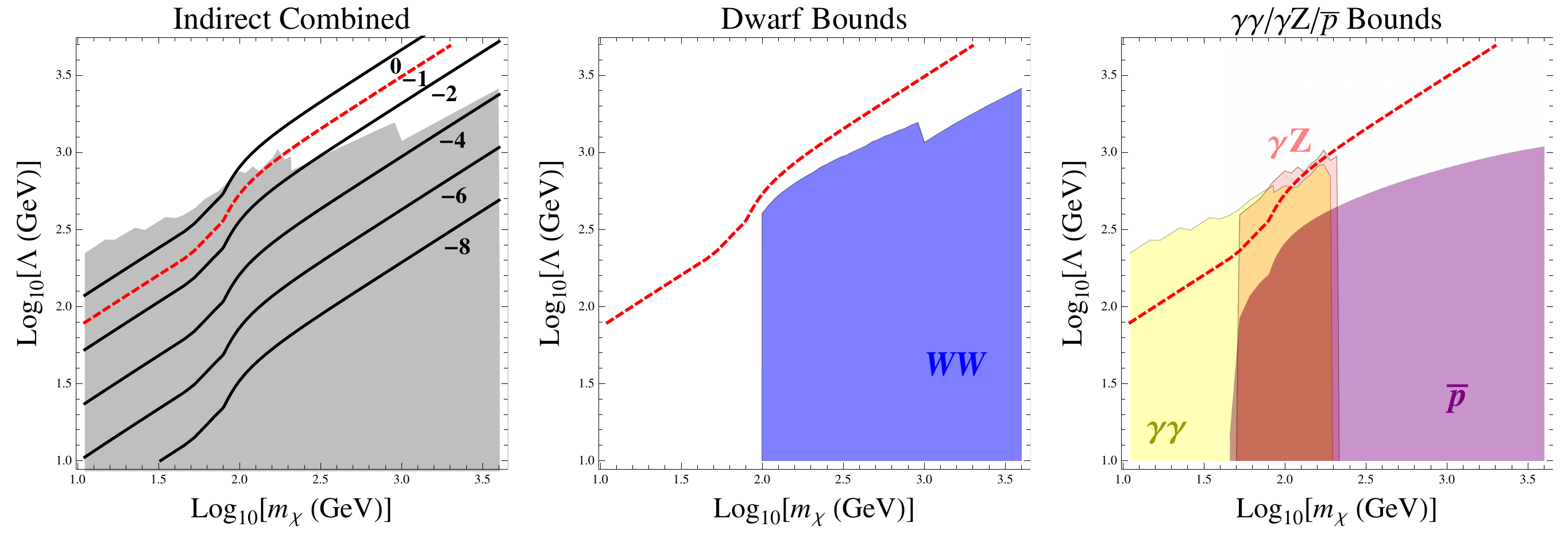}
    \caption{Same as in Figure \ref{figs:D5b} but for the operator \textbf{D7d}.}
    \label{figs:D7d}
  \end{figure}

 Figures \ref{figs:D5a}-\ref{figs:D7d} describe relic density contours and indirect detection bounds for our list of
operators in the $\Lambda$ vs.\ $m_{\chi}$ plane. The red-dashed line in all figures denotes where the thermal relic density
calculated for the given operator (in isolation) matches the cold dark matter density measured by WMAP. Black lines in the 
left panel of all figures denote lines of constant relic density and may be used to scale annihilation signal rates to those
expected in any given UV complete DM model. The region which is excluded by one or more searches combined is shaded in grey
in the left panel of all figures. Additional panels show the regions excluded by each particular search separately, $\eg$, 
dwarf bounds on annihilation to $\mu$-pairs or antiproton bounds from the PAMELA $\bar{p}/p$ measurement (of course, not
 all searches are relevant for any given operator). The excluded regions shown in Figs.\ \ref{figs:D5a}-\ref{figs:D7d} were
 calculated for the particular operator combinations quoted in Eq.\ \ref{combinations}. As the gauge bosons are now in the final state
for annihilation, there is no issue of coherence effects and it is fairly easy to derive excluded regions for
other subprocess combinations, given the results presented here.

Many of the features of Figs.\ \ref{figs:D5a}-\ref{figs:D7d} are easy to understand. The basic reach of the 
excluded regions varies quite a lot among the different operators according to velocity suppression intuition.
Operators D5a, D6b, D7a, D7d are all velocity suppressed to some extent (\emph{none} of the plotted region is
excluded for the Higgs portal operator D5a), and so are relatively difficult to exclude. Operators with 3-point
couplings to the $Z^0$ feature a prominent peak in sensitivity near $m_{\chi}\sim m_Z/2$ (while the relic
density contours simultaneously peak). The various searches have effective sensitivities within obvious bounds: annihilation to $WW$
only occurs for $m_{\chi}\geq m_W$, line searches are only effective for DM masses that generate lines of energy
between the \textit{Fermi}-LAT analysis thresholds\footnote{$E_{\gamma}=m_{\chi}$ for the $\gamma\gamma$ final state and 
$E_{\gamma}=(m_{\chi}-m_Z^2/4m_{\chi}$) for $\gamma Z$.} and antiproton bounds reach as far down in $m_{\chi}$ as there
are channels that produce hadronic matter ($WW$, $b\bar{b}$, $\tau\bar{\tau}$). Looking at the dwarf bounds, we
 see that excluded regions for $WW$ and $b\bar{b}$ final states feature a discontinuity around $m_{\chi}\sim 1\tev$, where the
 bound transitions from the \textit{Fermi}-LAT limit to the VERITAS limit\footnote{We also observe that
 there is no such discontinuity in the $\mu\bar{\mu}$ and $\tau\bar{\tau}$ excluded regions. The reason for this
 is that the \textit{Fermi}-LAT and VERITAS limits on stiff spectra (from $\mu\bar{\mu}$ and $\tau\bar{\tau}$
 annihilations) approximately match up at their overlap, whereas the
 limits on softer spectra ($WW$ and $b\bar{b}$) do not.}.

Recall that the excluded regions for all searches are calculated under the assumption that $\Omega h^2_{\chi}=\Omega h^2_{WMAP}$ and
 that annihilations in the current epoch occur only through our contact operators. We see that this scenario is excluded for operators
D5c and D6a for $m_{\chi}\lsim m_Z/2$, via the \textit{Fermi}-LAT dwarf $b\bar{b}$ limit, and for operators D7b-c for $m_{\chi}\lsim m_W$, via
the \textit{Fermi}-LAT $\gamma$ line limits. In the former case this is a model-independent statement (upon requiring that $\chi$
 is not milli-charged), and this exclusion complements the constraint $\Gamma_{Z\mathrm{,inv.}}\lsim 2\mev$ in this region (as we will see in the next section).
In the latter case this is a model-dependent
statement, as models that give D7b-c operators with negligible $\gamma\gamma$ annihilation are not excluded. In the case where excluded regions in
 Figs.\ \ref{figs:D5a}-\ref{figs:D7d} do not reach up to the red line, where a standard thermal cosmological calculation using
our operator would give $\Omega h^2_{\chi}=\Omega h^2_{WMAP}$, then the excluded scenarios have a \emph{larger} current
annihilation cross-section than what would give $\Omega h^2_{\chi}=\Omega h^2_{WMAP}$ in a thermal cosmology. Such scenarios
would require either \textit{(i)} non-thermal cosmological evolution or \textit{(ii)} inclusion of (relatively inefficient)
 annihilation channels other than our operators to boost the relic density of our $\chi$ up to the WMAP value.

\section{Combined Results}
\begin{figure}[hbtp]
  \centering
  \includegraphics[width=0.8\textwidth]{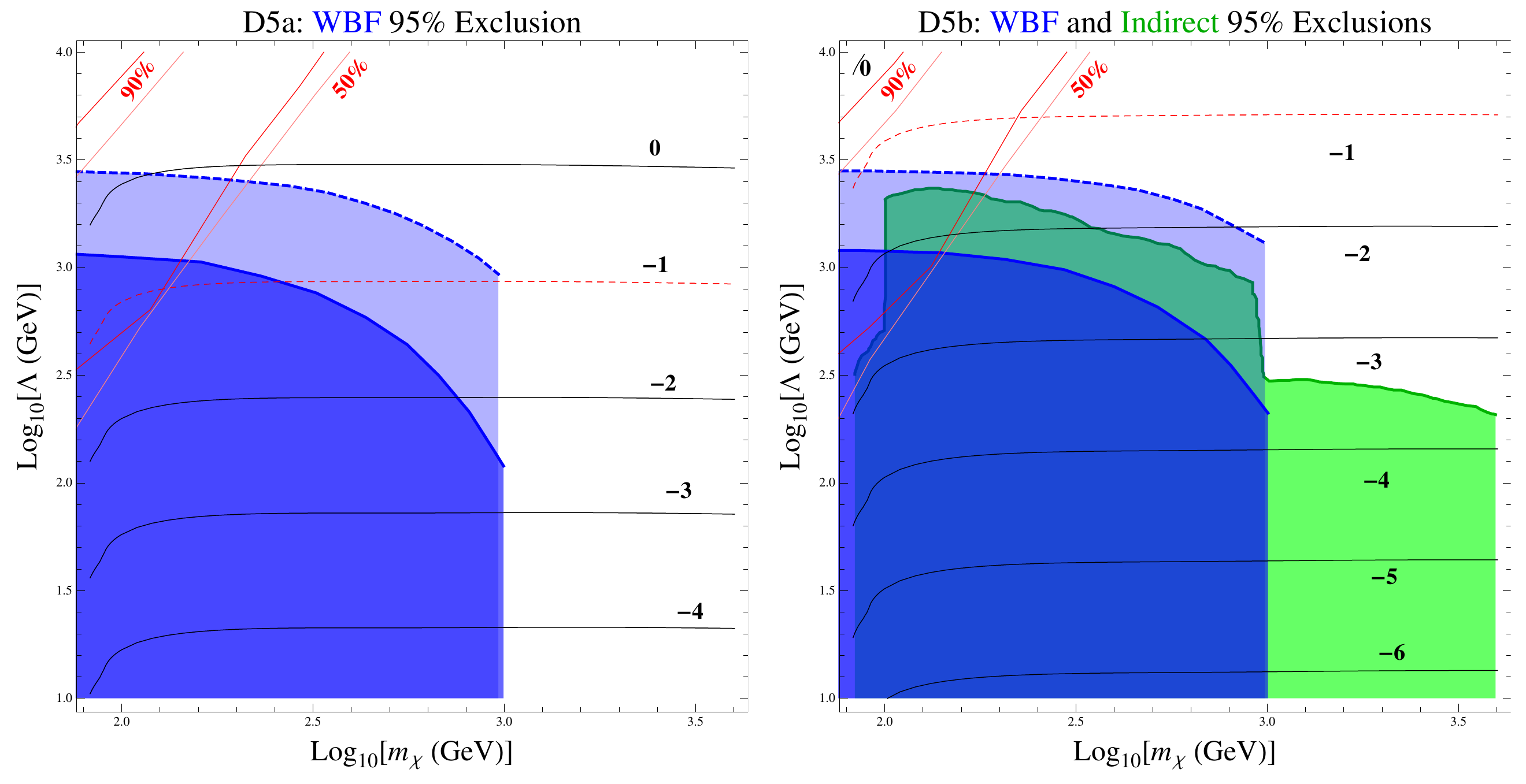}
  \caption{
    Regions excluded by $14\tev$ LHC WBF searches ($8\tev$- blue region, $14\tev$- light blue region) and current indirect detection searches (green region) are compared
    in the $\Lambda$ vs.\ $m_{\chi}$ plane for operators D5a (left panel) and D5b (right panel). Relic
    density is shown using the same contours that were described in Figs.\ \ref{figs:D5a}-\ref{figs:D7d}. 
    Unitarity curves as described before for Figs.\ \ref{figs:D5ab}-\ref{figs:D7ab} are reproduced here as well.
}
  \label{figs:comboD5ab}
\end{figure}

\begin{figure}[hbtp]
  \centering
  \includegraphics[width=0.8\textwidth]{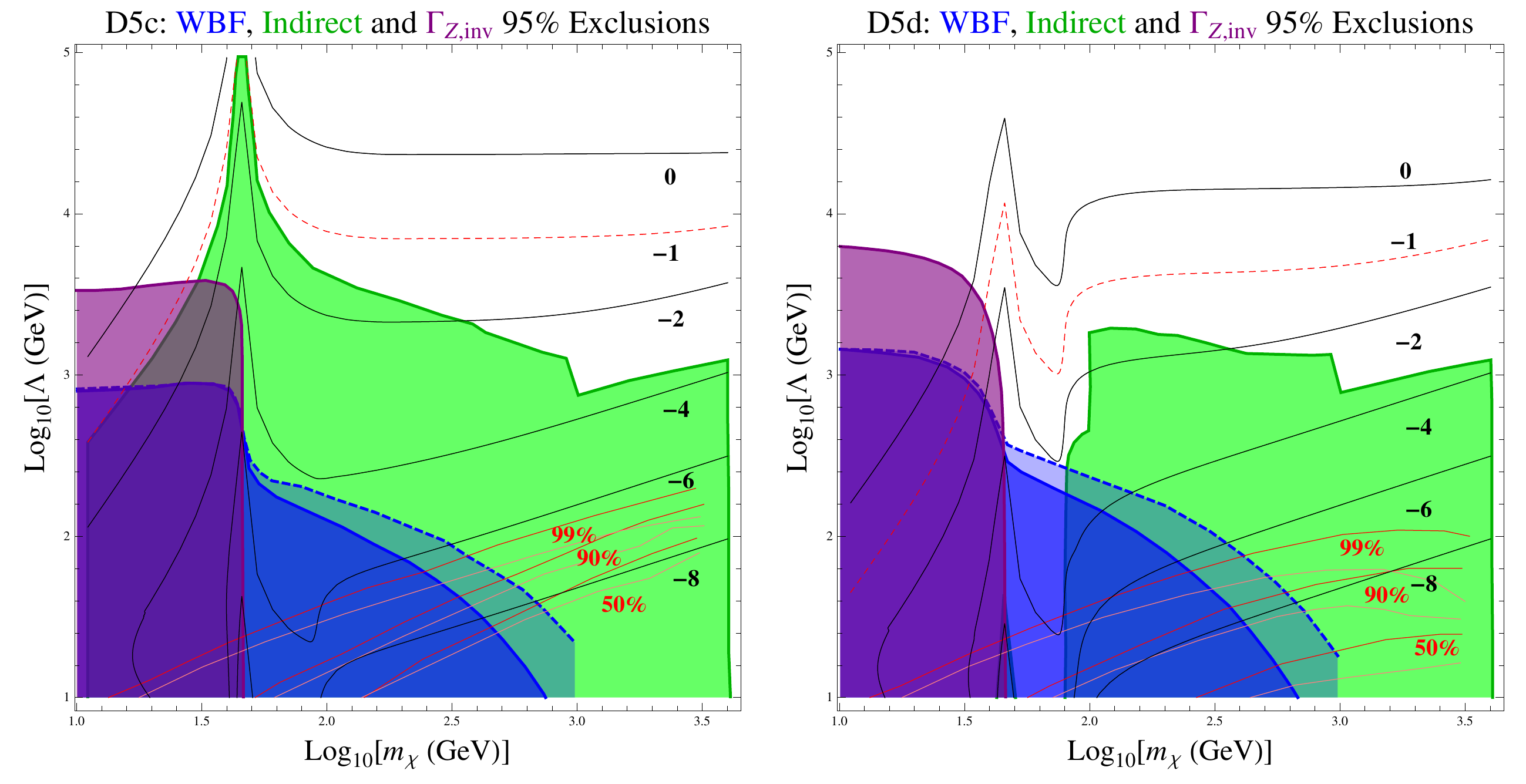}
  \caption{
    Regions excluded by $14\tev$ LHC WBF searches ($8\tev$- blue region, $14\tev$- light blue region), current indirect detection searches (green region) and measurements of the
    invisible width of the $Z^0$ (purple region) are compared in the $\Lambda$ vs.\ $m_{\chi}$ plane for operators D5c (left panel) and
    D5d (right panel). Relic density is shown using the same contours that were described in Figs.\ \ref{figs:D5a}-\ref{figs:D7d}.
    Unitarity curves as described before for Figs.\ \ref{figs:D5ab}-\ref{figs:D7ab} are reproduced here as well.
}
  \label{figs:comboD5cd}
\end{figure}

\begin{figure}[hbtp]
  \centering
  \includegraphics[width=0.8\textwidth]{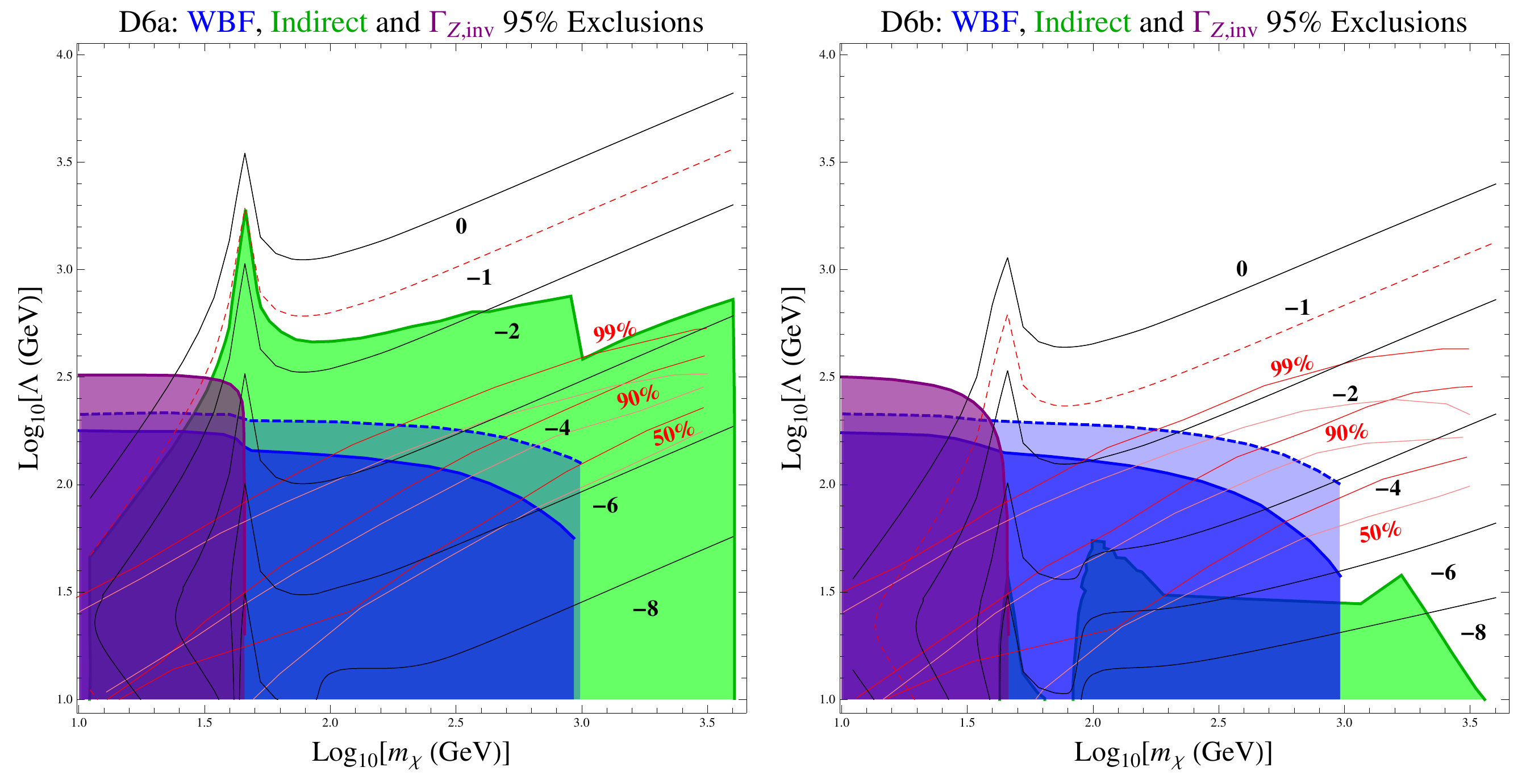}
  \caption{Same as in Figure \ref{figs:comboD5cd} but for operators D6a (left panel) and D6b (right panel).}
  \label{figs:comboD6ab}
\end{figure}

\begin{figure}[hbtp]
  \centering
  \includegraphics[width=0.8\textwidth]{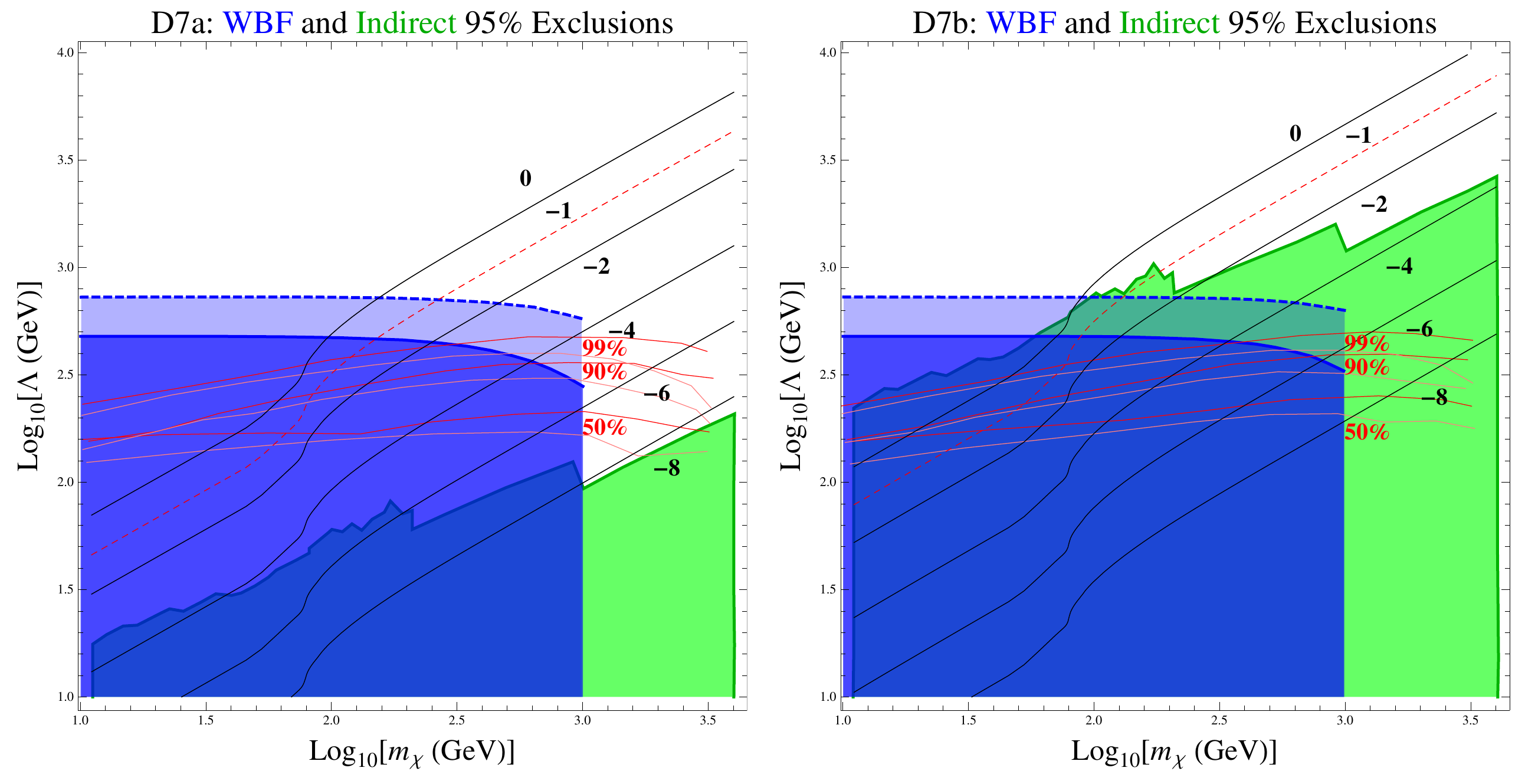}
  \caption{Same as in Figure \ref{figs:comboD5ab} but for operators D7a (left panel) and D7b (right panel).}
  \label{figs:comboD7ab}
\end{figure}

\begin{figure}[hbtp]
  \centering
  \includegraphics[width=0.8\textwidth]{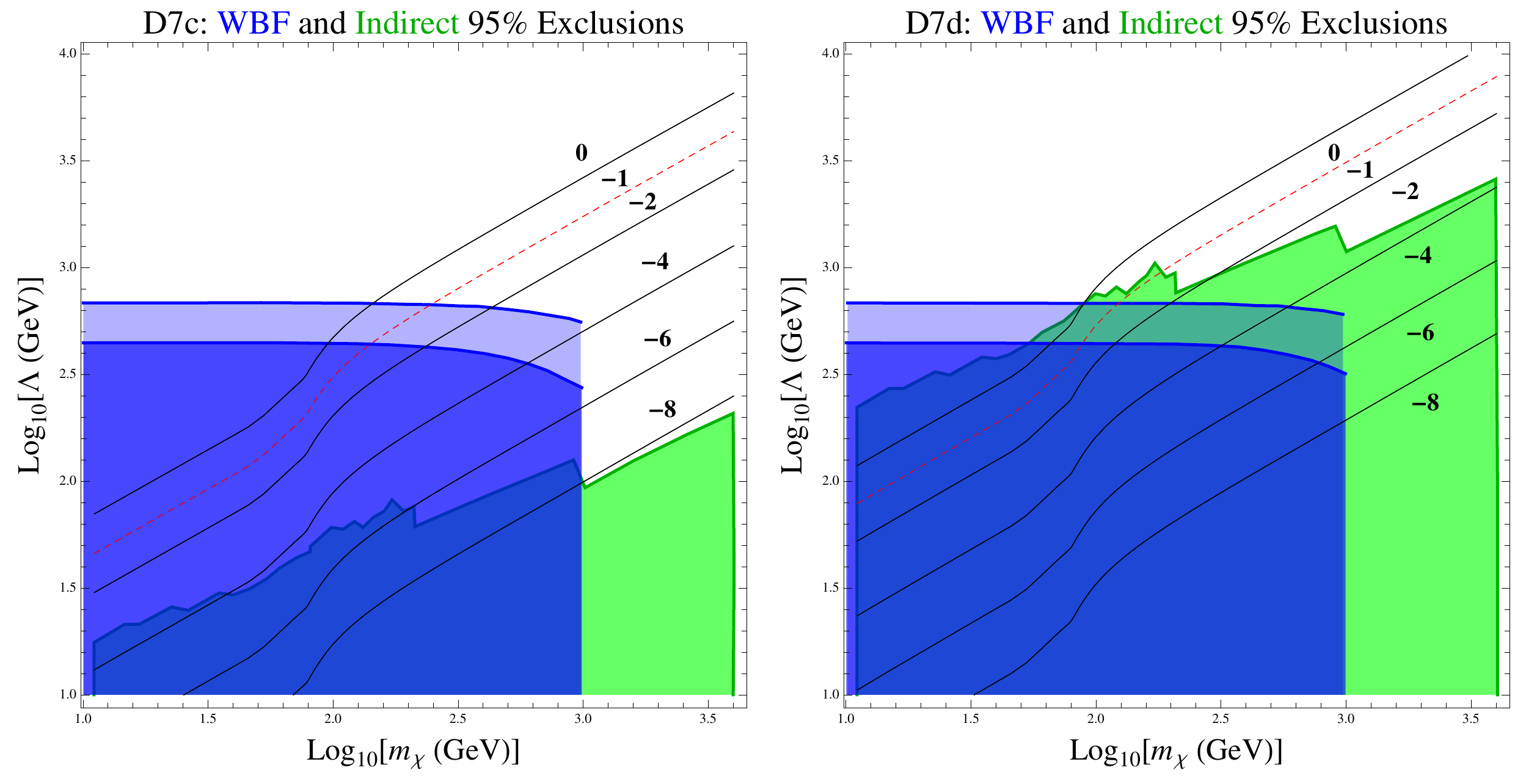}
  \caption{Same as in Figure \ref{figs:comboD5ab} but for operators D7c (left panel) and D7d (right panel). Unitarity curves were not
  calculated for these operators but are expected to be similar to those in Fig.\ \ref{figs:comboD7ab}.}
  \label{figs:comboD7cd}
\end{figure}

In Figures \ref{figs:comboD5ab}-\ref{figs:comboD7cd} we combine regions that can be excluded at 95\% C.L. by
 $8\tev$ $25 fb^{-1}$ and $14\tev$ $100 fb^{-1}$ weak boson fusion searches at the LHC, regions that are excluded at 95\% C.L.
 by current indirect detection searches and regions that are excluded by the $Z^0$ invisible width constraint, in the
 $\Lambda$ vs.\ $m_{\chi}$ plane. For most operators WBF searches reach $\Lambda\sim 500\gev-2\tev$ for DM masses $m_{\chi}\lsim 1\tev$, though 
operators that give three-point vertices are much harder to constrain in WBF, constraining $\Lambda\geq 100\gev$ only for DM masses below a 
couple of hundred GeV. Indirect detection reach varies wildly depending on the operator (predominantly according to velocity suppression, or lack thereof) 
and it is worth noting that these searches provide the only constraints for $m_{\chi}\gsim 1\tev$. For applicable operators, regions excluded by the
invisible width of the $Z^0$ provide the tightest constraints for $m_{\chi}<m_Z/2$. 

It is interesting to compare excluded regions to the overlaid relic density contours shown in Figs.\ \ref{figs:comboD5ab}-\ref{figs:comboD7cd},
and to discuss what this means for our assumptions that \emph{(i)} the DM abundance matches the WMAP value $\Omega h^2_{\chi}\approx 0.11$, \emph{(ii)} that
current annihilations proceed dominantly through our contact operators. Recall that the red-dashed lines in our Figures is the curve $\Lambda(m_{\chi})$
such that the thermal relic density, calculated using our operator in isolation, gives $\Omega h^2_{\chi}\approx 0.11$. For operators D5c, D5d, D6a and D6b,
the $Z^0$ invisible width constraint excludes such a scenario for $m_{\chi}\lsim m_Z/2$, implying that DM of this kind with $m_{\chi}\lsim m_Z/2$ necessarily
requires additional operators or dark sector annihilations in order to avoid overclosing the universe. WBF searches could similarly exclude this scenario
for operators D7a-d with $m_{\chi}\lsim 100-200\gev$. For heavier $m_{\chi}$ we can only exclude scenarios with relatively larger annihilation cross-sections
($\ie$, scenarios that would require non-thermal cosmology, or something similar, in order to supply $\Omega h^2_{\chi}\approx 0.11$).


\section{Discussion}

In this work we have investigated bounds on fermionic dark matter which interacts with the Standard Model primarily via electroweak gauge bosons, 
using a model-independent effective field theory description. In this picture we can easily compare the experimental reach 
of different classes of DM search experiments, here focusing on the only relevant searches (given our assumptions): Weak Boson Fusion events in $8\tev$
 or $14\tev$ LHC data, $\gamma$-ray observations of Milky Way dwarf spheroidals, $\gamma$-ray line searches, cosmic-ray antiproton data and constraints from 
the measured invisible width of the $Z^0$. 

The combined reach of these experiments is seen to probe the UV cutoff scale, $\Lambda$, up to weak-scale
values (hundreds of $\gev$ up to several $\tev$). Scenarios in which dark matter interacts with Standard Model
particles primarily via electroweak gauge bosons are, of course, harder to constrain than scenarios in
which the dominant interactions are with strongly interacting particles. Nevertheless, the finding that searches
are probing values of $\Lambda$ near the weak scale is interesting as these are natural values for the strength of such interactions
in the WIMP dark matter paradigm.

We have discussed the implications of these bounds for the possible cosmological evolution of such 
dark matter, finding that relatively light dark matter scenarios ($m_{\chi}\lsim m_Z/2$ or $m_{\chi}\lsim 100-200\gev$, depending on the operator) 
necessarily require additional structure (additional important operators or a non-trivial dark sector) to avoid overclosing the universe.

\section{Acknowledgments}
\label{sec:Acknowledgments}
  The authors would like to thank J. Alwall, Y.~Bai, N.~D.~Christensen, L. Dixon, C. Duhr, R. Harnik, M. Peskin and J. Wacker
  for discussions related to this work. This work was supported
 by the Department of Energy, Contract DE-AC02-76SF00515.

\end{document}